\documentclass[12pt]{article}
\usepackage[left=1in,right=1in,top=1in]{geometry}

\usepackage{amsmath,amssymb,amsthm}
\usepackage{bm}

\usepackage[T1]{fontenc}
\usepackage{times}

\usepackage{array,authblk,enumitem,hyperref,multirow,url,xspace}

\usepackage[linesnumbered,ruled,vlined]{algorithm2e}
\SetAlFnt{\small}

\usepackage{graphicx}
\usepackage{color}
\usepackage[skip=2pt]{caption}
\usepackage{subcaption}

\usepackage[sort,numbers]{natbib}

\newtheorem{example}{Example}
\newtheorem{lemma}{Lemma}
\newtheorem{theorem}{Theorem}

\DeclareMathOperator*{\argmax}{argmax}

\begin{document}
\title{Coresets for Minimum Enclosing Balls over Sliding Windows}

\author[1]{Yanhao Wang}
\author[2]{Yuchen Li}
\author[1]{Kian-Lee Tan}
\affil[1]{School of Computing, National University of Singapore, Singapore}
\affil[2]{School of Information Systems, Singapore Management University, Singapore}
\affil[ ]{$^1$\textit{\{yanhao90, tankl\}@comp.nus.edu.sg\quad $^2$yuchenli@smu.edu.sg}}
\maketitle

\begin{abstract}
\emph{Coresets} are important tools to generate concise summaries of massive datasets for approximate analysis.
A coreset is a small subset of points extracted from the original point set such that
certain geometric properties are preserved with provable guarantees.
This paper investigates the problem of maintaining a coreset to preserve the minimum enclosing ball (MEB)
for a sliding window of points that are continuously updated in a data stream.
Although the problem has been extensively studied in batch and append-only streaming settings,
no efficient sliding-window solution is available yet.
In this work, we first introduce an algorithm, called AOMEB, to build a coreset
for MEB in an append-only stream. AOMEB improves the practical performance of
the state-of-the-art algorithm while having the same approximation ratio.
Furthermore, using AOMEB as a building block, we propose two novel algorithms, namely
SWMEB and SWMEB+, to maintain coresets for MEB over the sliding window
with constant approximation ratios.
The proposed algorithms also support coresets for MEB in a reproducing kernel Hilbert space (RKHS).
Finally, extensive experiments on real-world and synthetic datasets demonstrate that
SWMEB and SWMEB+ achieve speedups of up to four orders of magnitude over the state-of-the-art
batch algorithm while providing coresets for MEB with rather small errors compared to the optimal ones.
\end{abstract}

\section{Introduction}\label{sec:introduction}
Unprecedented growth of data poses significant challenges in
designing algorithms that can scale to massive datasets.
Algorithms with superlinear complexity often become
infeasible on datasets with millions or billions of points.
\emph{Coresets} are effective approaches to tackling the challenges of big data analysis.
A coreset is a small subset extracted from the original point set
such that certain geometric properties are preserved with provable guarantees~\cite{DBLP:conf/kdd/BachemL018}.
Instead of processing the original dataset, one can perform the computation on its coreset with little loss of accuracy.
Various types of problems have been shown to be effective under coreset approximation,
e.g., $k$-median and $k$-means clustering~\cite{DBLP:conf/stoc/Har-PeledM04,DBLP:conf/kdd/BachemL018,DBLP:conf/soda/BravermanLLM16},
non-negative matrix factorization (NMF)~\cite{DBLP:conf/kdd/FeldmanT15},
kernel density estimation (KDE)~\cite{DBLP:conf/kdd/ZhengP17,DBLP:conf/compgeom/PhillipsT18},
and many others~\cite{DBLP:conf/kdd/BifetHPG11,DBLP:conf/nips/HugginsCB16,DBLP:conf/wsdm/CeccarelloPP18}.

Coresets for minimum enclosing balls
(MEB)~\cite{DBLP:conf/stoc/BadoiuHI02,DBLP:conf/cccg/Zarrabi-ZadehC06,DBLP:journals/jea/KumarMY03,DBLP:journals/comgeo/BadoiuC08,DBLP:conf/soda/AgarwalS10,DBLP:journals/siamjo/Yildirim08,DBLP:journals/cgf/LarssonK13}
have received significant attention due to its wide applications in
clustering~\cite{DBLP:conf/stoc/BadoiuHI02,DBLP:journals/jea/KumarMY03,DBLP:journals/comgeo/BadoiuC08},
support vector machines~\cite{DBLP:journals/jmlr/TsangKC05},
kernel regression~\cite{DBLP:conf/isnn/WeiL08}, fuzzy inference~\cite{DBLP:journals/tfs/ChungDW09},
shape fitting~\cite{DBLP:journals/cgf/LarssonK13},
and approximate furthest neighbor search~\cite{DBLP:conf/sisap/PaghSSS15}.
Given a set of points $P$, the minimum enclosing ball of $P$, denoted by $\mathsf{MEB}(P)$,
is the smallest ball that contains all points in $P$. A subset $S \subset P$
is a $\mu$-coreset for $\mathsf{MEB}(P)$ if the distance between the
center $\mathbf{c}^*(S)$ of $\mathsf{MEB}(S)$ and any point in $P$
is within $\mu \cdot r^*(S)$, where $r^*(S)$ is the radius of $\mathsf{MEB}(S)$~\cite{DBLP:journals/comgeo/BadoiuC08}.
Existing studies~\cite{DBLP:journals/jea/KumarMY03,DBLP:journals/comgeo/BadoiuC08} show that
there always exists a $(1+\varepsilon)$-coreset of size $O(\frac{1}{\varepsilon})$ ($0<\varepsilon<1$)
for the MEB of any point set, which is independent of the dataset size and dimension.

Most existing
methods~\cite{DBLP:conf/stoc/BadoiuHI02,DBLP:journals/jea/KumarMY03,DBLP:journals/comgeo/BadoiuC08,DBLP:journals/siamjo/Yildirim08,DBLP:journals/cgf/LarssonK13}
of coresets for MEB focus on the batch setting and
must keep all points in memory when constructing the coresets.
In many applications, such as network monitoring, financial analysis, and sensor data mining,
one needs to process data in the \emph{streaming} model~\cite{TCS-002}
where the input points arrive one at a time and cannot be stored entirely.
There have been several
methods~\cite{DBLP:journals/jacm/AgarwalHV04,DBLP:conf/cccg/Zarrabi-ZadehC06,DBLP:conf/soda/AgarwalS10,DBLP:journals/comgeo/Chan06}
to maintain coresets for MEB in data streams.
The state-of-the-art streaming algorithm~\cite{DBLP:conf/soda/AgarwalS10}
can maintain a $(\sqrt{2}+\varepsilon)$-coreset for MEB with a single
pass through the input dataset.
However, these algorithms only consider the append-only scenario where
new points are continuously added to, but old ones are never deleted from, the stream.
Hence, they fail to capture recency in time-sensitive applications
since the computation may be performed on outdated data.
To meet the recency requirement, the \emph{sliding window}
model~\cite{DBLP:journals/siamcomp/DatarGIM02,DBLP:conf/focs/BravermanO07,DBLP:conf/www/EpastoLVZ17,DBLP:conf/soda/BravermanLLM16}
that only considers the most recent $N$ points in the stream at any time
is a popular approach for real-time analytics.
One can trivially adapt append-only methods for the sliding window model but
a complete coreset reconstruction is deemed inevitable once an expired point is deleted.
To the best of our knowledge, there is no existing algorithm that
can maintain coresets for MEB over the sliding window efficiently.

\textbf{Our Results.}
In this paper, we investigate the problem of maintaining
coresets for MEB in the sliding window model. In particular, our results
are summarized as follows.
\begin{itemize}
  \item In Section~\ref{subsec:append:only}, we present the AOMEB algorithm to maintain
  a $(\sqrt{2}+\varepsilon)$-coreset of size $O(\frac{\log\theta}{\varepsilon^2})$
  with $O(\frac{m\log\theta}{\varepsilon^3})$ computation time per point for the MEB of an append-only stream,
  where $m$ is the dimension of the points and
  $\theta$ is the ratio of the maximum and minimum distances between any two points in the input dataset.
  AOMEB shows better empirical performance than the algorithm in~\cite{DBLP:conf/soda/AgarwalS10}
  while having the same approximation ratio.
  \item In Section~\ref{subsec:swmeb}, using AOMEB as a building block, we propose the SWMEB algorithm
  for coreset maintenance over a sliding window $W_t$ of the most recent $N$ points at time $t$.
  SWMEB divides $W_t$ into equal-length partitions.
  On each partition, it maintains a sequence of indices where each index corresponds to an instance of AOMEB.
  Theoretically, SWMEB can return a $(\sqrt{2}+\varepsilon)$-coreset for $\mathsf{MEB}(W_t)$
  with $O(\sqrt{N}\cdot\frac{m\log\theta}{\varepsilon^3})$ time
  and $O(\sqrt{N}\cdot\frac{\log\theta}{\varepsilon^2})$ space complexity, where $N$ is the window size.
  \item In Section~\ref{subsec:swmeb:plus}, we propose the SWMEB+ algorithm to improve upon SWMEB.
  SWMEB+ only maintains one sequence of indices, as well as the corresponding AOMEB instances, over $W_t$.
  By keeping fewer indices, SWMEB+ is more efficient than SWMEB in terms of time and space.
  Specifically, it only stores $O(\frac{\log^2 \theta}{\varepsilon^3})$ points with
  $O(\frac{m\log^2 \theta}{\varepsilon^4})$ processing time per point,
  both of which are independent of $N$.
  At the same time, it can still return a $(9.66+\varepsilon)$-coreset for $\mathsf{MEB}(W_t)$.
  \item In Section~\ref{subsec:kernel:meb}, we generalize our proposed algorithms to maintain coresets
  for MEB in a reproducing kernel Hilbert space (RKHS).
  \item In Section~\ref{sec:experiments}, we conduct extensive experiments on real-world and synthetic datasets
  to evaluate the performance of our proposed algorithms.
  The experimental results demonstrate that (1) AOMEB outperforms the state-of-the-art streaming
  algorithm~\cite{DBLP:conf/soda/AgarwalS10} in terms of coreset quality and efficiency;
  (2) SWMEB and SWMEB+ can return coresets for MEB with rather small errors (mostly within $1\%$),
  which are competitive with AOMEB and other streaming algorithms;
  (3) SWMEB and SWMEB+ achieve 2 to 4 orders of magnitude speedups over batch algorithms while
  running between 10 and 150 times faster than AOMEB;
  (4) SWMEB+ further improves the efficiency of SWMEB by up to 14 times while providing
  coresets with similar or even better quality.
\end{itemize}

\section{Preliminaries and Related Work}\label{sec:problem}

\textbf{Coresets for MEB.}
For two $m$-dimensional points $\mathbf{p}=(p_1,\ldots,p_m)$, $\mathbf{q}=(q_1,\ldots,q_m)$,
the Euclidean distance between $\mathbf{p}$ and $\mathbf{q}$ is denoted by $d(\mathbf{p},\mathbf{q})=\sqrt{\sum_{i=1}^{m}(p_i-q_i)^2}$.
An $m$-dimensional (closed) ball with center $\mathbf{c}$ and radius $r$ is defined as
$B(\mathbf{c},r) = \{ \mathbf{p} \in \mathbb{R}^{m} : d(\mathbf{c},\mathbf{p}) \leq r \}$.
We use $\mathbf{c}(B)$ and $r(B)$ to denote the center and radius of ball $B$.
The $\mu$-expansion of ball $B(\mathbf{c},r)$, denoted as $\mu \cdot B$, is a ball centered at $\mathbf{c}$
with radius $\mu \cdot r$, i.e., $\mu \cdot B = B(\mathbf{c},\mu \cdot r)$.

Given a set of $n$ points $P=\{\mathbf{p}_1,\ldots,\mathbf{p}_n\} \subset \mathbb{R}^{m}$,
the minimum enclosing ball of $P$, denoted as $\mathsf{MEB}(P)$,
is the smallest ball that contains all points in $P$.
The center and radius of $\mathsf{MEB}(P)$ are represented by $\mathbf{c}^{*}(P)$ and $r^{*}(P)$.
For a parameter $\mu > 1$, a ball $B$ is a $\mu$-approximate MEB of $P$
if $P \subset B$ and $r(B) \leq \mu \cdot r^{*}(P)$.
A subset $S \subset P$ is a $\mu$-coreset for $\mathsf{MEB}(P)$,
or $\mu$-$\mathsf{Coreset}(P)$ for brevity,
if $P \subset \mu \cdot \mathsf{MEB}(S)$.
Since $S \subseteq P$ and $r^{*}(S) \leq r^{*}(P)$,
$\mu \cdot \mathsf{MEB}(S)$ is always a $\mu$-approximate MEB of $P$.

\textbf{Sliding Window Model.}
This work focuses on
maintaining coresets for MEB in append-only streaming and sliding window settings.
For a sequence of (possibly infinite) points
$P=\langle \mathbf{p}_{1}, \mathbf{p}_{2}, \ldots \rangle$
arriving continuously as a data stream
where $\mathbf{p}_{t}$ is the $t$-th point, we first consider
the problem of maintaining
a $\mu$-$\mathsf{Coreset}(P_{t})$ for $P_t=\{\mathbf{p}_{1},\ldots,\mathbf{p}_{t}\}$
at any time $t$.

Furthermore, we consider the count-based sliding
window\footnote{In this paper, we focus on the count-based sliding window model.
But our proposed approaches can be trivially extended to the time-based sliding window
model~\cite{DBLP:journals/siamcomp/DatarGIM02}.}
on the stream $P$:
given a window size $N$, the sliding window~\cite{DBLP:journals/siamcomp/DatarGIM02}
$W_{t}$ at any time $t$ always contains
the latest $N$ points, i.e., $W_t=\{\mathbf{p}_{t'},\ldots,\mathbf{p}_{t}\}$
where $t'=\max(1,t-N+1)$.
We consider the problem of maintaining a $\mu$-$\mathsf{Coreset}(W_t)$
for $W_t$ at any time $t$.

\begin{algorithm}
    \SetKwInOut{Input}{Input}\SetKwInOut{Output}{Output}
    \Input{A set of points $P = \{ \mathbf{p}_{1}, \ldots, \mathbf{p}_{n} \}$, a parameter $\varepsilon \in (0,1)$}
    \Output{A coreset $S$ for $\mathsf{MEB}(P)$}
    $\mathbf{p}_{a} \gets \argmax_{\mathbf{p} \in P} d(\mathbf{p}_{1},\mathbf{p})$,
    $\mathbf{p}_{b} \gets \argmax_{\mathbf{p} \in P} d(\mathbf{p}_{a},\mathbf{p})$\;
    \label{line:offline:init:1}
    $S \gets \{\mathbf{p}_{a},\mathbf{p}_{b}\}$\;
    \label{line:offline:init:2}
    Initialize $B(\mathbf{c},r)$ with $\mathbf{c} \gets (\mathbf{p}_{a}+\mathbf{p}_{b})/2$,
    $r \gets d(\mathbf{p}_{a},\mathbf{p}_{b})/2$\;
    \label{line:offline:init:3}
    \While{$\exists \mathbf{p} \in P \setminus S : \mathbf{p} \notin (1+\varepsilon) \cdot B$
        \label{line:offline:iter:s}}
    {
        $\mathbf{q} \gets \argmax_{\mathbf{p} \in P \setminus S} d(\mathbf{c},\mathbf{p})$,
        $S \gets S \cup \{\mathbf{q}\}$\;
        Update $B$ such that $B(\mathbf{c},r)=\mathsf{MEB}(S)$\;
        \label{line:offline:iter:t}
    }
    \Return{$S$}\;
    \label{line:offline:return}
    \caption{CoreMEB~\cite{DBLP:journals/comgeo/BadoiuC08}}
    \label{alg:coreset:offline}
\end{algorithm}

\textbf{Related Work.}
We review the literature on MEB computation and coresets for MEB.
G\"artner~\cite{DBLP:conf/esa/Gartner99} and Fischer et al.~\cite{DBLP:conf/esa/FischerGK03}
propose two pivoting algorithms that resemble the simplex method of linear programming
for computing exact MEBs. Both algorithms have an exponential complexity
w.r.t. the dimension $m$ and thus are not scalable for large datasets with high dimensions.
Subsequently, a line of research
work~\cite{DBLP:conf/stoc/BadoiuHI02,DBLP:journals/jea/KumarMY03,DBLP:journals/comgeo/BadoiuC08,DBLP:journals/siamjo/Yildirim08,DBLP:journals/cgf/LarssonK13}
studies the problem of building coresets to approximate MEBs.
They propose efficient batch algorithms for constructing
a $(1+\varepsilon)$-$\mathsf{Coreset}(P)$ of any point set $P$.
The basic scheme used in these algorithms is presented in Algorithm~\ref{alg:coreset:offline}.
First of all, it selects the point $\mathbf{p}_{a}$ furthest from $\mathbf{p}_{1}$
and $\mathbf{p}_{b}$ furthest from $\mathbf{p}_{a}$ out of $P$,
using $S=\{\mathbf{p}_{a},\mathbf{p}_{b}\}$ as the initial coreset
(Lines~\ref{line:offline:init:1} \&~\ref{line:offline:init:2}).
The center $\mathbf{c}$ and radius $r$ of $\mathsf{MEB}(S)$
can be computed from $\mathbf{p}_{a}$ and $\mathbf{p}_{b}$ directly
(Line~\ref{line:offline:init:3}).
Then, it iteratively picks the point $\mathbf{q}$ furthest from the current center $\mathbf{c}$,
adds $\mathbf{q}$ to $S$, and updates $B(\mathbf{c},r)$
so that $B$ is $\mathsf{MEB}(S)$, until no point in $P$
is outside of the $(1+\varepsilon)$-expansion of $B$
(Lines~\ref{line:offline:iter:s}--\ref{line:offline:iter:t}).
Finally, it returns $S$ as a coreset for $\mathsf{MEB}(P)$
(Line~\ref{line:offline:return}).
Theoretically, Algorithm~\ref{alg:coreset:offline} terminates in $O(\frac{1}{\varepsilon})$ iterations
and returns a $(1+\varepsilon)$-$\mathsf{Coreset}(P)$ of
size $O(\frac{1}{\varepsilon})$~\cite{DBLP:journals/comgeo/BadoiuC08}.
Compared with exact MEB solvers~\cite{DBLP:conf/esa/Gartner99,DBLP:conf/esa/FischerGK03},
coreset-based approaches
run in linear time w.r.t. the dataset size $n$ and dimension $m$,
and achieve better performance on high-dimensional data.
Nevertheless, they must store all points in memory
and process them in multiple passes, which are not suitable for data stream applications.

Several methods are proposed to approximate MEBs or
coresets for MEB in streaming and dynamic settings.
Agarwal et al.~\cite{DBLP:journals/jacm/AgarwalHV04} and Chan~\cite{DBLP:journals/comgeo/Chan06}
propose algorithms to build $(1+\varepsilon)$-coresets for MEB in append-only streams.
Though working well in low dimensions, both algorithms become impractical for higher dimensions (i.e., $m>10$)
due to $O(1/\varepsilon^{O(m)})$ complexity.
Zarrabi-Zadeh and Chan~\cite{DBLP:conf/cccg/Zarrabi-ZadehC06} propose
a $1.5$-approximate algorithm to compute MEBs in append-only streams.
Agarwal and Sharathkumar~\cite{DBLP:conf/soda/AgarwalS10}
design a data structure that can maintain $(\sqrt{2}+\varepsilon)$-coresets for MEB
and $(1.37+\varepsilon)$-approximate MEBs over append-only streams.
Chan and Pathak~\cite{DBLP:journals/comgeo/ChanP14} propose a method
for maintaining $(1.22+\varepsilon)$-approximate MEBs in the dynamic setting,
which supports the insertions and deletions of random points.
To the best of our knowledge, none of the existing methods can maintain
coresets for MEB over the sliding window efficiently.
All of them have to store the entire window of points and
recompute from scratch for every window slide, which is expensive in terms of time and space.

\section{Our Algorithms}\label{sec:algorithms}

In this section we present our algorithms to maintain coresets for MEB.
We first introduce a $(\sqrt{2}+\varepsilon)$-approximate append-only streaming algorithm,
called AOMEB, in Section~\ref{subsec:append:only}.
Using AOMEB as a building block,
we propose the SWMEB algorithm with the same $(\sqrt{2}+\varepsilon)$-approximation
ratio in Section~\ref{subsec:swmeb}.
Furthermore, we propose a more efficient SWMEB+ algorithm
that retains a constant approximation ratio in Section~\ref{subsec:swmeb:plus}.

\subsection{The AOMEB Algorithm}\label{subsec:append:only}

The AOMEB algorithm is inspired by CoreMEB~\cite{DBLP:journals/comgeo/BadoiuC08}
(see Algorithm~\ref{alg:coreset:offline}) to work in the append-only streaming model.
Compared with CoreMEB, which can access the entire dataset and \emph{optimally}
select the furthest point into the coreset at each iteration,
AOMEB is restricted to process the dataset in a single pass
and determine whether to include a point into the coreset or discard it
immediately after seeing it. Therefore, AOMEB adopts a \emph{greedy}
strategy for coreset maintenance: adding a new point to the coreset
once it is outside of the MEB w.r.t. the current coreset.

\begin{algorithm}
    \SetKwInOut{Input}{Input}\SetKwInOut{Output}{Output}
    \Input{A set of points $P = \{ \mathbf{p}_1, \ldots, \mathbf{p}_n \}$,
        a parameter $\varepsilon_1 \in (0,1)$}
    \Output{A coreset $S$ for $\mathsf{MEB}(P)$}
    $S_{1} \gets \{\mathbf{p}_1\}$ and
    initialize $B_1(\mathbf{c}_1,r_1)$ with $\mathbf{c}_1 \gets \mathbf{p}_1,r_1 \gets 0$\;
    \label{line:append:only:init}
    \For{$t \gets 2, \ldots, n$}
    {
        \uIf{$\mathbf{p}_t \notin (1+\varepsilon_1) \cdot B_{t-1}$
            \label{line:append:only:streaming:s}}
        {
            $S_t \gets S_{t-1} \cup \{ \mathbf{p}_t \}$\;
            Update $B_{t-1}$ to $B_t(\mathbf{c}_t,r_t) = \mathsf{MEB}(S_t)$\;
            \label{line:append:only:update}
        }
        \Else
        {
            $S_t \gets S_{t-1}$ and $B_t \gets B_{t-1}$\;
            \label{line:append:only:streaming:t}
        }
    }
    \Return{$S \gets S_n$}\;
    \label{line:append:only:return}
    \caption{AOMEB}
    \label{alg:coreset:append:only}
\end{algorithm}

The pseudo code of AOMEB is presented in Algorithm~\ref{alg:coreset:append:only}.
First of all, it takes $S_{1}=\{\mathbf{p}_{1}\}$ as the initial coreset
with $B_1(\mathbf{p}_{1},0)$ as $\mathsf{MEB}(S_{1})$ (Line~\ref{line:append:only:init}).
Then, it performs a one-pass scan over the point set, using the procedure
in Lines~\ref{line:append:only:streaming:s}--\ref{line:append:only:streaming:t}
for each point $\mathbf{p}_{t}$:
It first computes the distance between $\mathbf{p}_{t}$ and $\mathbf{c}_{t-1}$.
If $\mathbf{p}_{t} \in (1+\varepsilon_{1}) \cdot B_{t-1}$, no update is needed;
otherwise, it adds $\mathbf{p}_{t}$ to the coreset $S_{t-1}$ and updates $B_{t-1}$
to $\mathsf{MEB}(S_{t})$.
Finally, after processing all points in $P$, it returns $S_{n}$ as the coreset $S$
for $\mathsf{MEB}(P)$ (Line~\ref{line:append:only:return}).

\textbf{Theoretical Analysis.}
Next, we provide an analysis of the approximation ratio and complexity of AOMEB.
It is noted that the \emph{greedy} strategy of AOMEB is also adopted by existing streaming algorithms,
i.e., SSMEB~\cite{DBLP:conf/cccg/Zarrabi-ZadehC06} and blurred ball cover (BBC)~\cite{DBLP:conf/soda/AgarwalS10}.
Nevertheless, the update procedure is different:
SSMEB uses a simple geometric method to enlarge the MEB
such that both the previous MEB and the new point are contained
while AOMEB and BBC recompute the MEB once the coreset is updated.
As a result, AOMEB and BBC are less efficient than SSMEB but ensure a better approximation ratio.
Compared with BBC, which keeps the ``archives'' of MEBs for previous coresets,
AOMEB only maintains one MEB w.r.t.~$S_t$ at time $t$.
Therefore, AOMEB is more efficient than BBC in practice.
Next, we will prove that AOMEB has the same $(\sqrt{2}+\varepsilon)$-approximation as BBC.
First of all, we present the \emph{hemisphere property}~\cite{DBLP:conf/stoc/BadoiuHI02}
that forms the basis of our analysis.

\begin{lemma}[Hemisphere Property~\cite{DBLP:conf/stoc/BadoiuHI02}]\label{lm:hemisphere}
  For a set of points $P \subset \mathbb{R}^{m}$,
  any closed half-space that contains $\mathbf{c}^{*}(P)$ must contain at least
  a point $\mathbf{p} \in P$ such that $d(\mathbf{c}^{*}(P),\mathbf{p})=r^{*}(P)$.
\end{lemma}

The proof of Lemma~\ref{lm:hemisphere} can be found
in Section 2 of~\cite{DBLP:conf/stoc/BadoiuHI02}.
Based on Lemma~\ref{lm:hemisphere}, we can analyze the complexity and approximation ratio of AOMEB theoretically.

\begin{theorem}\label{thm:append:only:space}
  For any $\mathbf{p}_{t} \in S_{n}$, it holds that $r_{t} \geq (1+\frac{\varepsilon_1^2}{8})r_{t-1}$.
\end{theorem}
\begin{proof}
    If $\mathbf{p}_{t} \in S_{n}$, then $d(\mathbf{p}_{t},\mathbf{c}_{t-1}) > (1+\varepsilon_{1})r_{t-1}$.
    We discuss two cases of $d(\mathbf{c}_{t},\mathbf{c}_{t-1})$ separately.
    If $d(\mathbf{c}_{t},\mathbf{c}_{t-1}) \leq \frac{\varepsilon_1}{2}r_{t-1}$, then
    \begin{equation*}
    r_{t} \geq d(\mathbf{c}_{t},\mathbf{p}_{t})
    \geq d(\mathbf{p}_{t},\mathbf{c}_{t-1}) - d(\mathbf{c}_{t},\mathbf{c}_{t-1})
    \geq (1+\frac{\varepsilon_{1}}{2})r_{t-1}
    \end{equation*}
    If $d(\mathbf{c}_{t},\mathbf{c}_{t-1}) > \frac{\varepsilon_1}{2}r_{t-1}$, then let
    $H$ be a hyperplane passing through $\mathbf{c}_{t-1}$ with $\mathbf{c}_{t}-\mathbf{c}_{t-1}$ as its normal.
    Let $H^{+}$ be the closed half-space, bounded by $H$, that does not contain $\mathbf{c}_{t}$.
    According to Lemma~\ref{lm:hemisphere}, there must exist a point $\mathbf{q} \in S_{t-1} \cap H^{+}$
    such that $d(\mathbf{c}_{t-1},\mathbf{q})=r_{t-1}$. Thus,
    \begin{equation*}
    r_{t} \geq d(\mathbf{c}_{t},\mathbf{q})
    \geq \sqrt{d^{2}(\mathbf{c}_{t-1},\mathbf{q}) + d^2(\mathbf{c}_{t},\mathbf{c}_{t-1})}
    > \sqrt{1 + (\frac{\varepsilon_1}{2})^{2}} \cdot r_{t-1}
    \end{equation*}
    In addition, $1+\frac{\varepsilon_{1}}{2} > \sqrt{1 + (\frac{\varepsilon_1}{2})^{2}} > 1+\frac{\varepsilon_{1}^{2}}{8}$
    as $\varepsilon_{1} \in (0,1)$.
    Therefore, we prove that $r_{t} \geq (1+\frac{\varepsilon_1^2}{8})r_{t-1}$ in both cases.
\end{proof}

\begin{theorem}\label{thm:append:only:approx}
  For any $\mathbf{p}_{t} \in P$, it holds that $\mathbf{p}_{t} \in (\sqrt{2} + \varepsilon_{1}) \cdot B_n$.
\end{theorem}
\begin{proof}
    For any $\mathbf{p}_{t} \in P$, we have either $\mathbf{p}_{t} \in S_n$
    or $d(\mathbf{p}_{t},\mathbf{c}_{t}) \leq (1+\varepsilon_1)r_t$.
    If $\mathbf{p}_{t} \in S_n$, it is obvious that $\mathbf{p}_{t} \in (1 + \varepsilon_{1}) \cdot B_n$.
    If $\mathbf{p}_{t} \notin S_n$, we have $d(\mathbf{p}_{t},\mathbf{c}_{n})
    \leq d(\mathbf{c}_{t},\mathbf{c}_{n}) + (1+\varepsilon_1)r_t$.
    According to Lemma~\ref{lm:hemisphere}, there must exist a point $\mathbf{q} \in S_{t}$
    such that $d^2(\mathbf{c}_{n},\mathbf{q}) \geq d^2(\mathbf{c}_{t},\mathbf{c}_{n}) + r_t^2$.
    Therefore,
    \begin{multline*}
    d(\mathbf{p}_t,\mathbf{c}_n) \leq d(\mathbf{c}_{t},\mathbf{c}_{n}) + (1+\varepsilon_1)r_t
    \leq (d(\mathbf{c}_{t},\mathbf{c}_{n}) + r_t) +\varepsilon_1 r_t \\
    \leq \sqrt{2} \cdot \sqrt{d^2(\mathbf{c}_{t},\mathbf{c}_{n}) + r_t^2} + \varepsilon_1 r_t
    \leq \sqrt{2} \cdot d(\mathbf{c}_{n},\mathbf{q}) + \varepsilon_1 r_t \leq (\sqrt{2}+\varepsilon_1)r_n
    \end{multline*}
    We conclude that $\mathbf{p}_{t} \in (\sqrt{2} + \varepsilon_{1}) \cdot B_n, \forall \mathbf{p}_{t} \in P$.
\end{proof}

Theorem~\ref{thm:append:only:approx} indicates that AOMEB returns
a $(\sqrt{2} + \varepsilon)$-$\mathsf{Coreset}(P)$ where $\varepsilon = O(\varepsilon_1)$ for an arbitrary point set $P$.
According to Theorem~\ref{thm:append:only:space}, the radius of $\mathsf{MEB}(S_t)$
increases by $1+O(\varepsilon^{2})$ times whenever a new point is added to $S_t$.
After processing $\mathbf{p}_1$ and $\mathbf{p}_2$,
the coreset $S_2$ contains both points with $r_2 \geq d_{min}/2$
where $d_{min}=\min_{\mathbf{p},\mathbf{q} \in P \wedge \mathbf{p} \neq \mathbf{q}}
d(\mathbf{p},\mathbf{q})$.
In addition, the radius $r_n$ of $S_n$ is bounded by $r^{*}(P) < d_{max}$
where $d_{max}=\max_{\mathbf{p},\mathbf{q} \in P} d(\mathbf{p},\mathbf{q})$.
Therefore, $r_n/r_2 = O(\theta)$ where $\theta=d_{max}/d_{min}$
and the size of $S$ is $O(\frac{\log \theta}{\varepsilon^2})$.
Finally, the update procedure for each point $\mathbf{p}_t$ spends
$O(m)$ time to compute $d(\mathbf{c}_{t-1},\mathbf{p}_t)$
and $O(\frac{m\log \theta}{\varepsilon^3})$ time to update $B_t$.

\subsection{The SWMEB Algorithm}\label{subsec:swmeb}

In this subsection, we present the SWMEB algorithm for coreset maintenance over the sliding window $W_t$.
The basic idea is to adapt AOMEB for the sliding window model
by keeping multiple AOMEB instances with different starting points over $W_t$.
However, the key problem is to identify the appropriate indices, i.e., starting points, for these instances.
A naive scheme, i.e., creating a set of indices that are evenly distributed over $W_t$, cannot
give any approximation guarantee of coreset quality. Therefore, we design a partition-based scheme
for index maintenance in SWMEB: dividing $W_t$ into equal-length partitions and keeping a sequence
of indices on each partition such that at least one instance can provide an approximate coreset for $\mathsf{MEB}(W_t)$
at any time $t$.

\begin{figure}[t]
  \centering
  \includegraphics[width=0.75\textwidth]{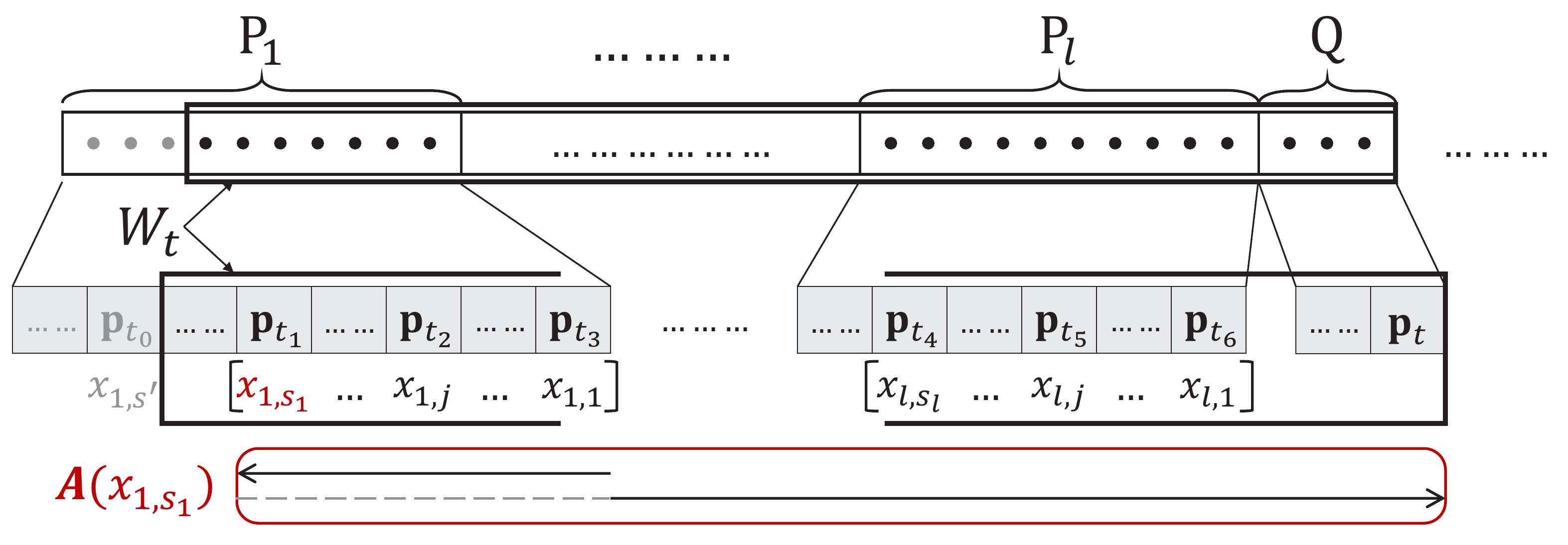}
  \caption{An illustration of the SWMEB algorithm. Two arrows indicate
  the order in which the points in $W_t$ are processed by $\mathcal{A}(x_{1,s_1})$.}
  \label{fig:swmeb}
\end{figure}

The procedure of SWMEB is illustrated in Figure~\ref{fig:swmeb}.
It divides $W_t$ into $l=N/L$ partitions $\{P_1,\ldots,P_l\}$ of equal length $L$.
It keeps a sequence of $s_i$ indices $\langle x_{i,1},\ldots,x_{i,s_i} \rangle$
from the end to the beginning of each partition $P_i$.
As $W_t$ slides over time, old points in $P_1$ expire (colored in \emph{grey})
while new points are temporarily stored in a buffer $Q$.
The index $x_{1,s_1}$ on $P_1$, which is the closest to the beginning of $W_t$, will be deleted once it expires.
When the size of $Q$ reaches $L$, it will delete $P_1$ and shift remaining partitions
as all points in $P_1$ must have expired.
Then, it creates a new partition $P_l$ for the points in $Q$
and the indices on $P_l$.
Moreover, each index $x_{i,j}$ corresponds to an AOMEB instance $\mathcal{A}(x_{i,j})$
that processes $P[x_{i,j},t]=\{\mathbf{p}_{x_{i,j}},\ldots,\mathbf{p}_t\}$ at any time $t$.
Specifically, $\mathcal{A}(x_{i,j})$ will process the points from the end of $P_i$ to $\mathbf{p}_{x_{i,j}}$
when $x_{i,j}$ is created and then update for each point till $\mathbf{p}_t$.
Finally, the coreset is always provided by $\mathcal{A}(x_{1,s_1})$.

\begin{algorithm}[t]
  \SetKwInOut{Input}{Input}\SetKwInOut{Output}{Output}
  \Input{A sequence of points $P = \langle \mathbf{p}_{1}, \mathbf{p}_{2}, \ldots \rangle$,
	     the window size $N$, the partition size $L$, two parameters $\varepsilon_{1},\varepsilon_{2} \in (0,1)$}
  \Output{A coreset $S_t$ for $\mathsf{MEB}(W_t)$}
  Initialize $l \gets 0, Q \gets \varnothing, X_0 \gets \varnothing$\;\label{line:swmeb:init}
  \For{$t \gets 1,2,\ldots$}
  {
    $Q \gets Q \cup \{\mathbf{p}_t\}$, $X_t \gets X_{t-1}$\;
    \label{line:swmeb:streaming:s}
    \label{line:swmeb:phase:1:s}
    \If{$|Q| = L$}
    {
      \uIf{$t \leq N$}
      {
        $l \gets l+1$, create a new partition $P_l \gets Q$\;
      }
      \Else
      {
        Drop $P_1$, shift $P_i$ (as well as the indices on $P_i$) to $P_{i-1}$ for $i\in[2,l]$,
        and create a new partition $P_l \gets Q$\;
      }
      \label{line:swmeb:phase:1:t}
      Initialize an instance $\mathcal{A}$ of Algorithm~\ref{alg:coreset:append:only},
      $r_l \gets 0$, $s_l \gets 0$\;
      \label{line:swmeb:phase:2:s}
      \For{$t' \gets t,\ldots,t-L+1$\label{line:swmeb:index:s}}
      {
        $\mathcal{A}$ processes $\mathbf{p}_{t'}$ with
        Line~\ref{line:append:only:streaming:s}--\ref{line:append:only:streaming:t}
        of Algorithm~\ref{alg:coreset:append:only} and maintains
        a coreset $S[t,t']$ and its MEB $B[t,t']$\;\label{line:swmeb:phase:1:c}
        \If{$r[t,t'] \geq (1+\varepsilon_2)r_l$\label{line:swmeb:index:cond}}
        {
          $r_l \gets r[t,t'], s_l \gets s_l+1$\;
          $x_{l,s_l} \gets t', X_t \gets X_t \cup \{x_{l,s_l}\}$\;
          $\mathcal{A}(x_{l,s_l}) \gets \mathcal{A}$ after processing $\mathbf{p}_{t'}$\;
        }
        \label{line:swmeb:index:t}
      }
      $Q,P_l \gets \varnothing$\;
    }
    \label{line:swmeb:phase:2:t}
    \If{$x_{1,s_1} < t-N+1$ \label{line:swmeb:phase:3:s}}
    {
      $X_t \gets X_{t} \setminus \langle x_{1,s_1} \rangle$, terminate $\mathcal{A}(x_{1,s_1})$, and $s_1 \gets s_1-1$\;
    }
    \label{line:swmeb:phase:3:t}
    \For{$i \gets 1,\ldots,l$ \textbf{and} $j \gets 1,\ldots,s_i$\label{line:swmeb:phase:4:s}}
    {
      $\mathcal{A}(x_{i,j})$ processes $\mathbf{p}_t$ with
      Line~\ref{line:append:only:streaming:s}--\ref{line:append:only:streaming:t}
      of Algorithm~\ref{alg:coreset:append:only} and maintains
      a coreset $S[x_{i,j},t]$ and its MEB $B[x_{i,j},t]$\;
    }
    \label{line:swmeb:phase:4:t}
    \label{line:swmeb:streaming:t}
    return $S_t \gets S[x_{1,s_1},t]$ as the coreset for $\mathsf{MEB}(W_t)$\;
    \label{line:swmeb:return}
  }
  \caption{SWMEB}\label{alg:swmeb}
\end{algorithm}

The pseudo code of SWMEB is presented in Algorithm~\ref{alg:swmeb}.
For initialization, the latest partition ID $l$ is set to $0$
and the buffer $Q$ as well as the indices $X_0$ are set to $\varnothing$ (Line~\ref{line:swmeb:init}).
Then, it processes all points in the stream $P$ one by one with the procedure of
Lines~\ref{line:swmeb:streaming:s}--\ref{line:swmeb:streaming:t},
which can be separated into four phases as follows.
\begin{itemize}
  \item \textbf{Phase 1 (Lines~\ref{line:swmeb:phase:1:s}--\ref{line:swmeb:phase:1:t}):}
  After adding a new point $\mathbf{p}_t$ to $Q$,
  it checks the size of $Q$.
  If $|Q|=L$, a new partition will be created for $Q$.
  When $t \leq N$, it increases $l$ by $1$ and creates a new partition $P_{l}$.
  Otherwise, $P_1$ must have expired and thus is dropped.
  Then, the partitions $\{ P_2,\ldots,P_l \}$ (and the indices on each partition)
  are shifted to $\{ P_1,\ldots,P_{l-1} \}$
  and a new partition $P_{l}$ is created.
  \item \textbf{Phase 2 (Lines~\ref{line:swmeb:phase:2:s}--\ref{line:swmeb:phase:2:t}):}
  Next, it creates the indices and corresponding AOMEB instances on $P_l$.
  It runs an AOMEB instance $\mathcal{A}$ to process each point in $P_l$ inversely from $\mathbf{p}_t$ to $\mathbf{p}_{t-L+1}$.
  Initially, the number of indices $s_l$ on $P_l$ and the radius $r_l$ w.r.t. the latest index $x_{l,s_l}$ are $0$.
  We denote the coreset maintained by $\mathcal{A}$ after processing $\mathbf{p}_{t'}$
  as $S[t,t']$. Then, $\mathsf{MEB}(S[t,t'])$ is represented by $B[t,t']$ with radius $r[t,t']$.
  If $r[t,t'] \geq (1+\varepsilon_2)r_l$, it will update $r_l$ to $r[t,t']$,
  add a new index $x_{l,s_l}$ to $X_t$, and
  use the snapshot of $\mathcal{A}$ after processing $\mathbf{p}_{t'}$ as $\mathcal{A}(x_{l,s_l})$.
  After the indices on $P_l$ is created, $Q$ will be reset for new incoming points.
  \item \textbf{Phase 3 (Lines~\ref{line:swmeb:phase:3:s}--\ref{line:swmeb:phase:3:t}):}
  It checks whether $x_{1,s_1}$, i.e., the earliest index on $P_1$, has expired.
  If so, it will delete $x_{1,s_1}$ from $X_t$ and terminate $\mathcal{A}(x_{1,s_1})$ accordingly.
  \item \textbf{Phase 4 (Lines~\ref{line:swmeb:phase:4:s}--\ref{line:swmeb:phase:4:t}):}
  For each index $x_{i,j} \in X_t$ with $i \in [1,l]$ and $j \in [1,s_i]$ at time $t$,
  it updates the corresponding AOMEB instance $\mathcal{A}(x_{i,j})$ by processing $\mathbf{p}_t$.
\end{itemize}
Finally, it always returns $S[x_{1,s_1},t]$ from $\mathcal{A}(x_{1,s_1})$ as
the coreset $S_t$ for $\mathsf{MEB}(W_t)$ at time $t$ (Line~\ref{line:swmeb:return}).

\textbf{Theoretical Analysis.}
In the following, we will first prove the approximation ratio of $S_t$ returned by SWMEB for $\mathsf{MEB}(W_t)$.
Then, we discuss the time and space complexity of SWMEB.

We first prove the following lemma that will be used in subsequent analyses.
\begin{lemma}\label{lm:meb:properties}
    For any two point sets $P_1,P_2$ such that $P_2 \subset P_1$,
    it must hold that $d^2(\mathbf{c}^{*}(P_1),\mathbf{c}^{*}(P_2)) \leq (r^{*}(P_1))^2 - (r^{*}(P_2))^2$.
\end{lemma}
\begin{proof}
    Obviously, Lemma~\ref{lm:meb:properties} must hold when $\mathbf{c}^{*}(P_1)=\mathbf{c}^{*}(P_2)$.
    When $\mathbf{c}^{*}(P_1) \neq \mathbf{c}^{*}(P_2)$, we consider a hyperplane $H$ passing
    through $\mathbf{c}^{*}(P_2)$ with $\mathbf{c}^{*}(P_1)-\mathbf{c}^{*}(P_2)$ as its normal.
    Let $H^{+}$ be the close half-space, bounded by $H$, that does not contain $\mathbf{c}^{*}(P_1)$.
    According to Lemma~\ref{lm:hemisphere},
    there must exist $\mathbf{q} \in P_2 \cap H^{+}$ such that $d(\mathbf{c}^{*}(P_2),\mathbf{q})=r^{*}(P_2)$.
    In addition, $d(\mathbf{c}^{*}(P_1),\mathbf{q}) \leq r^{*}(P_1)$ for $\mathbf{q} \in P_1$.
    Finally, as $\mathbf{c}^{*}(P_1) \notin H^{+}$, $\mathbf{q} \in H^{+}$,
    and $\mathbf{c}^{*}(P_1)-\mathbf{c}^{*}(P_2)$ is the normal of $H$,
    we acquire $d^2(\mathbf{c}^{*}(P_1),\mathbf{q}) \geq d^2(\mathbf{c}^{*}(P_2),\mathbf{q})
    + d^2(\mathbf{c}^{*}(P_1),\mathbf{c}^{*}(P_2))$.
    Thus, it holds that $d^2(\mathbf{c}^{*}(P_1),\mathbf{c}^{*}(P_2)) \leq (r^{*}(P_1))^{2} - (r^{*}(P_2))^{2}$
    and we conclude the proof.
\end{proof}

\begin{theorem}\label{thm:swmeb:approx}
  For any $\mathbf{p} \in W_t$, it holds that $\mathbf{p} \in (\sqrt{2}+\varepsilon) \cdot \mathsf{MEB}(S_t)$
  where $\varepsilon=O(\varepsilon_1+\sqrt{\varepsilon_2})$.
\end{theorem}
\begin{proof}
    According to Algorithm~\ref{alg:swmeb}, the instance $\mathcal{A}(x_{1,s_1})$ is always used to return
    the coreset $S_t$. Since $\mathcal{A}(x_{1,s_1})$ has already processed the points
    from $\mathbf{p}_{x_{1,s_1}}$ to $\mathbf{p}_t$, these points must be contained in
    $(\sqrt{2}+\varepsilon_1) \cdot \mathsf{MEB}(S_t)$ according to Theorem~\ref{thm:append:only:approx}.
    Thus, we only need to consider the points from $\mathbf{p}_{t-N+1}$ to $\mathbf{p}_{x_{1,s_1}-1}$.
    To create the indices on $P_1$, we process the points of $P_1$ with
    an AOMEB instance (see Line~\ref{line:swmeb:phase:1:c}).
    Here we use $S_1,B_1(\mathbf{c}_1,r_1)$ and $S_2,B_2(\mathbf{c}_2,r_2)$ to
    denote the coresets of this instance and the corresponding MEBs
    after processing $\mathbf{p}_{x_{1,s_1}}$ and $\mathbf{p}_{t'}$ ($t-N<t'<x_{1,s_1}$) respectively.
    For each point $\mathbf{p}_{t'}$, it holds that
    $\mathbf{p}_{t'} \in (1+\varepsilon_1) \cdot B_2$ and $r_2 < (1+\varepsilon_2) r_1$.
    In addition, as $S_1 \subseteq S_2$, we have $d^2(\mathbf{c}_1,\mathbf{c}_2) \leq r^2_2 - r^2_1$
    from Lemma~\ref{lm:meb:properties}.
    Therefore,
    \begin{multline*}
    d(\mathbf{c}_1,\mathbf{p}_{t'}) \leq d(\mathbf{c}_1,\mathbf{c}_2) + d(\mathbf{c}_2,\mathbf{p}_{t'})
    \leq (1+\varepsilon_1)r_2 + \sqrt{r^2_2 - r^2_1}
    \\ \leq (1+\varepsilon_1)(1+\varepsilon_2)r_1 + \sqrt{(1+\varepsilon_2)^2-1}\cdot r_1
    = \big(1+O(\varepsilon_1+\sqrt{\varepsilon_2})\big)r_1
    \end{multline*}
    In addition, according to Lemma~\ref{lm:hemisphere}, there must exist a point $\mathbf{q} \in S_1$
    such that $d^2(\mathbf{c}_t,\mathbf{q}) \geq d^2(\mathbf{c}_1,\mathbf{c}_t) + r_1^2$.
    Let $\varepsilon=O(\varepsilon_1+\sqrt{\varepsilon_2})$, we have
    \begin{multline*}
    d(\mathbf{p}_{t'},\mathbf{c}_t) \leq d(\mathbf{c}_1,\mathbf{c}_t) + (1+\varepsilon)r_1
    \leq \big(d(\mathbf{c}_1,\mathbf{c}_t)+r_1 \big) + \varepsilon r_1
    \\ \leq \sqrt{2} \cdot \sqrt{d^2(\mathbf{c}_1,\mathbf{c}_t) + r_1^2} + \varepsilon r_1
    \leq \sqrt{2} \cdot d(\mathbf{c}_t,\mathbf{q}) + \varepsilon r_1 \leq (\sqrt{2}+\varepsilon)r_t
    \end{multline*}
    We prove that $\mathbf{p}_{t'} \in (\sqrt{2} + \varepsilon) \cdot B_t$
    and thus conclude the proof.
\end{proof}

Theorem~\ref{thm:swmeb:approx} shows that $S_t$ returned by SWMEB
is a $(\sqrt{2}+\varepsilon)$-$\mathsf{Coreset}(W_t)$
where $\varepsilon=O(\varepsilon_1+\sqrt{\varepsilon_2})$ at any time $t$.
To analyze the complexity of SWMEB, we first consider the number of indices in $X_t$.
For each partition, SWMEB maintains $O(\frac{\log \theta}{\varepsilon_2})$ indices
where $\theta=d_{max}/d_{min}$.
Thus, $X_t$ contains $O(\frac{l \log \theta}{\varepsilon_2})$ indices
and the number of points stored by SWMEB is $O(\frac{l \log^2 \theta}{\varepsilon_1^2\varepsilon_2} + L)$.
Furthermore, the time of SWMEB to update a point $\mathbf{p}_t$ comprises
(1) the time to maintain the instance w.r.t. each index in $X_t$ for $\mathbf{p}_t$ and
(2) the amortized time to create the indices for each partition.
Overall, the time complexity of SWMEB to update each point is $O(\frac{lm \log^2 \theta}{\varepsilon_1^3\varepsilon_2})$.
As $l=N/L$, the number of points maintained by SWMEB is minimal
when $L=\frac{\log \theta}{\varepsilon_1}\cdot\sqrt{\frac{N}{\varepsilon_2}}$.
In this case, the number of points stored by SWMEB is $O(\sqrt{N} \cdot \frac{\log\theta}{\varepsilon^2})$
and the time complexity of SWMEB to update one point is $O(\sqrt{N} \cdot \frac{m\log\theta}{\varepsilon^3})$.

\subsection{The SWMEB+ Algorithm}\label{subsec:swmeb:plus}

\begin{figure}[t]
  \centering
  \includegraphics[width=0.75\textwidth]{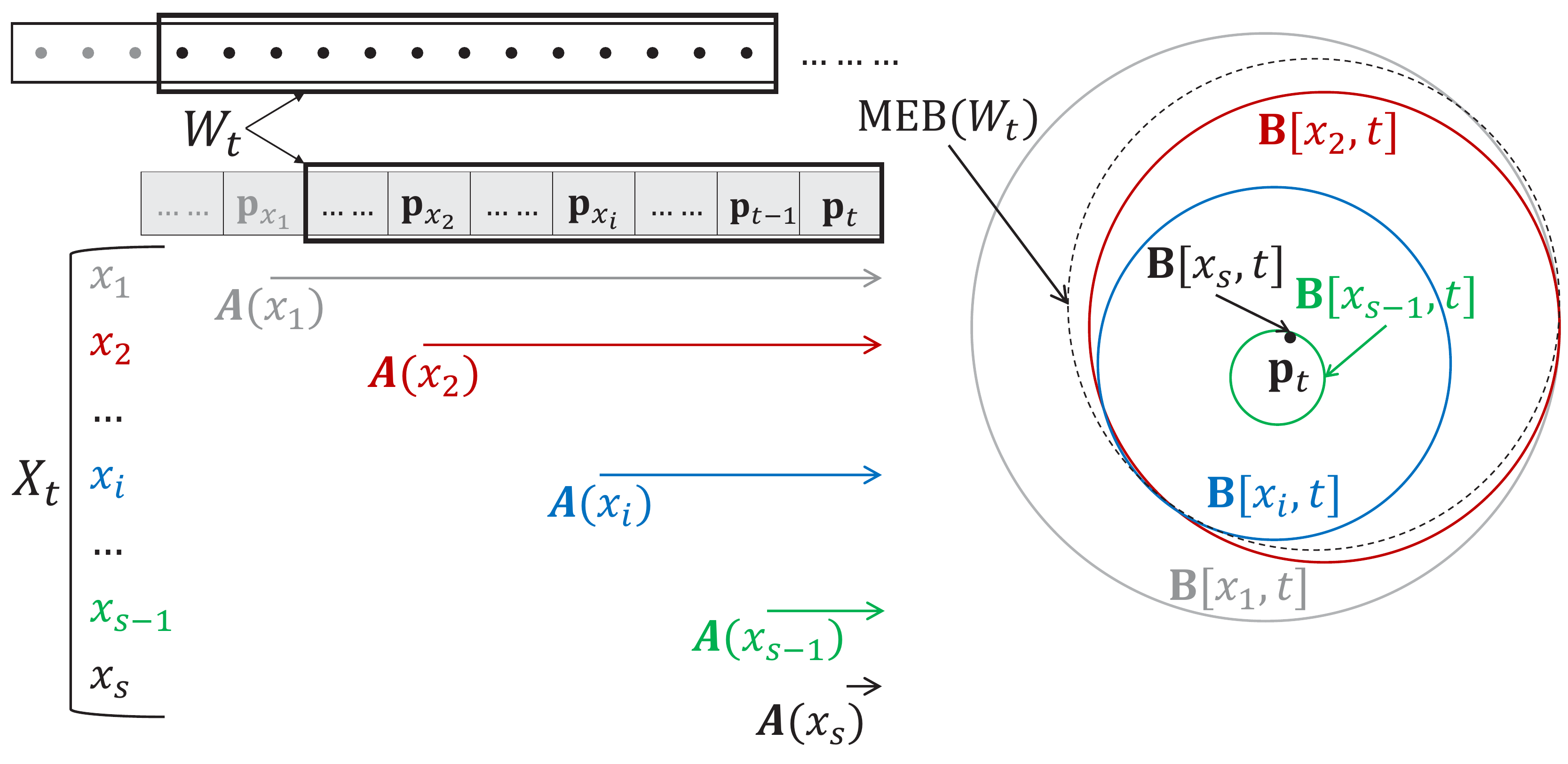}
  \caption{An illustration of the SWMEB+ algorithm.}
  \label{fig:swmeb:plus}
\end{figure}

\begin{algorithm}[t]
  \SetKwInOut{Input}{Input}\SetKwInOut{Output}{Output}
  \Input{A sequence of points $P = \langle \mathbf{p}_{1}, \mathbf{p}_{2}, \ldots \rangle$,
	     the window size $N$, two parameters $\varepsilon_{1},\varepsilon_{2} \in (0,1)$}
  \Output{A coreset $S_t$ for $\mathsf{MEB}(W_t)$}
  Initialize $s \gets 0, X_{0} \gets \varnothing$\;\label{line:swmebp:init}
  \For{$t \gets 1,2,\ldots$}
  {
    $s \gets s+1, x_s \gets t$, and $X_{t} \gets X_{t-1} \cup \langle x_s \rangle$\;
    \label{line:swmebp:streaming:s}
    \label{line:swmebp:phase:1:s}
    Initialize an instance $\mathcal{A}(x_s)$ of Algorithm~\ref{alg:coreset:append:only}\;
    \label{line:swmebp:phase:1:t}
    \While{$x_2 < t-N+1$\label{line:swmebp:phase:2:s}}
    {
      $X_t \gets X_t \setminus \langle x_1 \rangle$, terminate $\mathcal{A}(x_1)$\;
      Shift the remaining indices in $X_t$, $s \gets s-1$\;
      \label{line:swmebp:phase:2:t}
    }
    \For{$i \gets 1, \ldots, s$\label{line:swmebp:phase:3:s}}
    {
      $\mathcal{A}(x_i)$ processes $\mathbf{p}_{t}$ with
      Line~\ref{line:append:only:streaming:s}--\ref{line:append:only:streaming:t}
      of Algorithm~\ref{alg:coreset:append:only}, maintaining
      a coreset $S[x_i,t]$ and its MEB $B[x_i,t]$\;
      \label{line:swmebp:phase:3:t}
    }
    \While{$\exists i : 1 \leq i \leq s-2 \wedge r[x_i,t] \leq (1+\varepsilon_2) r[x_{i+2},t]$\label{line:swmebp:phase:4:s}}
    {
      $X_{t} \gets X_{t} \setminus \langle x_{i+1} \rangle$,
      terminate $\mathcal{A}(x_{i+1})$\;
      Shift the remaining indices in $X_t$, $s \gets s-1$\;
      \label{line:swmebp:phase:4:t}
      \label{line:swmebp:streaming:t}
    }
    \uIf{$x_1 \geq t-N+1$\label{line:swmebp:return:s}}
    {
      return $S_t \gets S[x_1,t]$ as the coreset for $\mathsf{MEB}(W_t)$\;
    }
    \Else
    {
      return $S_t \gets S[x_2,t]$ as the coreset for $\mathsf{MEB}(W_t)$\;
      \label{line:swmebp:return:t}
    }
  }
  \caption{SWMEB+}\label{alg:swmeb:plus}
\end{algorithm}

In this subsection we present the SWMEB+ algorithm that improves upon SWMEB
in terms of time and space while still achieving a constant approximation ratio.
The basic idea of SWMEB+ is illustrated in Figure~\ref{fig:swmeb:plus}.
Different from SWMEB, SWMEB+ only maintains a single sequence of $s$ indices $X_t=\{x_1,\ldots,x_s\}$ over $W_t$.
Then, each index $x_i$ also corresponds to an AOMEB instance $\mathcal{A}(x_i)$
that processes a substream of points from $\mathbf{p}_{x_i}$ to $\mathbf{p}_{t}$.
We use $S[x_i,t]$ for the coreset returned by $\mathcal{A}(x_i)$ at time $t$
and $B[x_i,t]$ centered at $\mathbf{c}[x_i,t]$ with radius $r[x_i,t]$ for $\mathsf{MEB}(S[x_i,t])$.
Furthermore, SWMEB+ maintains the indices based on the radii of the MEBs.
Specifically, given any $\varepsilon_2>0$, for three neighboring indices $x_{i},x_{i+1},x_{i+2}$,
if $r[x_i,t] \leq (1+\varepsilon_2) r[x_{i+2},t]$,
then $x_{i+2}$ is considered as a good approximation for $x_{i}$ and thus $x_{i+1}$ can be deleted.
In this way, the radii of the MEBs gradually decreases from $x_1$ to $x_s$,
with the ratios of any two neighboring indices close to $(1+\varepsilon_2)$.
Any window starting between $x_i$ and $x_{i+1}$ is approximated by $\mathcal{A}(x_{i+1})$.
Finally, SWMEB+ keeps at most one expired index (and must be $x_1$)
in $X_t$ to track the upper bound for the radius $r^{*}(W_t)$ of $\mathsf{MEB}(W_t)$.
The AOMEB instance corresponding to
the first non-expired index ($x_1$ or $x_2$) provides the coreset for $\mathsf{MEB}(W_t)$.

The pseudo code of SWMEB+ is presented in Algorithm~\ref{alg:swmeb:plus}.
In the initialization phase, $X_0$ and $s$ are set to $\varnothing$ and $0$ respectively
(Line~\ref{line:swmebp:init}).
Then, all points in $P$ are processed one by one with the procedure of
Lines~\ref{line:swmebp:streaming:s}--\ref{line:swmebp:streaming:t},
which includes four phases as follows.
\begin{itemize}
  \item \textbf{Phase 1 (Lines~\ref{line:swmebp:phase:1:s}--\ref{line:swmebp:phase:1:t}):}
  Upon the arrival of $\mathbf{p}_{t}$ at time $t$, it creates a new index $x_s=t$ and adds $x_s$ to $X_t$;
  accordingly, an AOMEB instance $\mathcal{A}(x_s)$ w.r.t.~$x_s$
  is initialized to process the substream beginning at $\mathbf{p}_t$.
  \item \textbf{Phase 2 (Lines~\ref{line:swmebp:phase:2:s}--\ref{line:swmebp:phase:2:t}):}
  When there exists more than one expired index (i.e., earlier than the beginning of $W_t$),
  it deletes the first index $x_1$ and terminates $\mathcal{A}(x_1)$ until there is only one expired index.
  Note that it shifts the remaining indices after deletion to always guarantee
  $x_i$ is the $i$-th index of $X_t$.
  \item \textbf{Phase 3 (Lines~\ref{line:swmebp:phase:3:s}--\ref{line:swmebp:phase:3:t}):}
  For each $x_i \in X_t$, it updates the instance $\mathcal{A}(x_i)$ for $\mathbf{p}_t$.
  The update procedure follows Line~\ref{line:append:only:streaming:s}--\ref{line:append:only:streaming:t}
  of Algorithm~\ref{alg:coreset:append:only}. After the update, $\mathcal{A}(x_i)$
  maintains a coreset $S[x_i,t]$ and its MEB $B[x_i,t]$
  by processing a stream $P[x_i,t]=\langle \mathbf{p}_{x_i},\ldots,\mathbf{p}_{t} \rangle$.
  \item \textbf{Phase 4 (Lines~\ref{line:swmebp:phase:4:s}--\ref{line:swmebp:phase:4:t}):}
  It executes a scan of $X_t$ from $x_1$ to $x_t$ to delete the indices that
  can be approximated by their successors.
  For each $x_i \in X_t$ ($i \leq s-2$), it checks the radii $r[x_i,t]$
  and $r[x_{i+2},t]$ of $B[x_i,t]$ and $B[x_{i+2},t]$.
  If $r[x_i,t] \leq (1+\varepsilon_2) r[x_{i+2},t]$,
  then it deletes the index $x_{i+1}$ from $X_t$, terminates $\mathcal{A}(x_{i+1})$,
  and shifts the remaining indices accordingly.
\end{itemize}
After performing the above procedure, it returns either $S[x_1,t]$
(when $x_1$ has not expired) or $S[x_2,t]$ (when $x_1$ has expired) as the coreset $S_t$
for $\mathsf{MEB}(W_t)$ at time $t$.

\begin{figure}[t]
  \centering
  \includegraphics[width=0.25\textwidth]{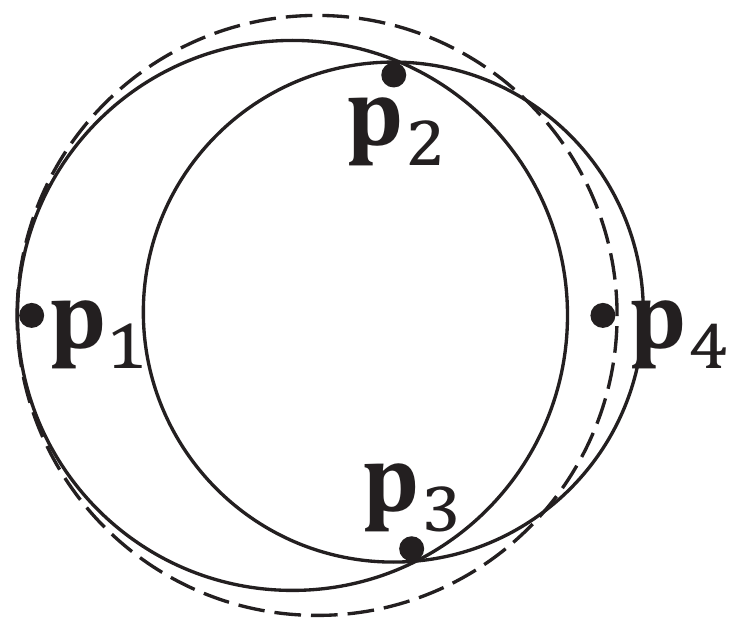}
  \caption{An illustration of Example~\ref{counter:example}.}
  \label{fig:counter:example}
\end{figure}

\textbf{Theoretical Analysis.}
The strategy of index maintenance based on the ratios of radii is inspired by
\emph{Smooth Histograms}~\cite{DBLP:conf/focs/BravermanO07} for estimating
stream statistics over sliding windows.
However, \emph{Smooth Histograms} cannot be applied to our problem
because it requires an oracle to provide a $(1+\varepsilon)$-approximate
function value in any append-only stream~\cite{DBLP:conf/www/EpastoLVZ17}
but any practical solution (i.e.,~\cite{DBLP:conf/soda/AgarwalS10} and AOMEB) only gives a
$(\sqrt{2}+\varepsilon)$-approximation for $r^*(P)$ of an append-only stream $P$.
In addition, \emph{Smooth Histograms} are also used for submodular maximization
in the sliding window model~\cite{DBLP:conf/www/EpastoLVZ17,DBLP:journals/pvldb/WangFLT17,8410031}.
Nevertheless, such an extension is still not applicable for our problem because
the radius function $r^*(\cdot)$ is not submodular in the view of set functions,
which is shown by Example~\ref{counter:example}.
In the following, we will prove that SWMEB+ still has a constant approximation ratio
by an analysis that is different from~\cite{DBLP:conf/focs/BravermanO07,DBLP:conf/www/EpastoLVZ17}.

\begin{example}\label{counter:example}
    A function $f(\cdot)$ is submodular if $f(P \cup \{\mathbf{p}\}) - f(P) \geq f(Q \cup \{\mathbf{p}\}) - f(Q)$
    for any set $P \subseteq Q$ and point $\mathbf{p} \notin Q$.
    In Figure~\ref{fig:counter:example}, for $P_1=\{\mathbf{p}_1,\mathbf{p}_2,\mathbf{p}_3\}$
    and $P_2=\{\mathbf{p}_2,\mathbf{p}_3\}$, $r^*(P_1 \cup \{\mathbf{p}_4\}) - r^*(P_1) > 0$
    but $r^*(P_2 \cup \{\mathbf{p}_4\}) - r^*(P_2) = 0$.
    As $P_2 \subset P_1$, we show that $r^*(\cdot)$ is not a submodular function.
\end{example}

We first present two lemmas that will be used in our analysis of SWMEB+.
\begin{lemma}\label{lm:swmebp:1}
    For a set of points $P$, let $B^{*}(\mathbf{c}^{*},r^{*})=\mathsf{MEB}(P)$
    and $B(\mathbf{c},r)=\mathsf{MEB}(S)$ where $S \subseteq P$ is
    the coreset returned by Algorithm~\ref{alg:coreset:append:only}.
    If $\frac{r^{*}}{r}=\mu$ ($1 \leq \mu \leq \sqrt{2} + \varepsilon_1$),
    then $B^{*} \subset (\mu + \sqrt{\mu^2 - 1}) \cdot B$.
\end{lemma}
\begin{proof}
    According to Lemma~\ref{lm:meb:properties},
    $d^{2}(\mathbf{c}^{*},\mathbf{c}) \leq (r^{*})^{2} - r^2$ because of $S \subseteq P$.

    Then $\forall \mathbf{p}^{*} \in B^{*}$, it holds that
    \begin{equation*}
    d(\mathbf{c},\mathbf{p}^{*}) \leq d(\mathbf{c}^{*},\mathbf{c}) + d(\mathbf{c}^{*},\mathbf{p}^{*})
    \leq \sqrt{(r^{*})^{2} - r^2} + r^{*} = (\mu + \sqrt{\mu^2 - 1})r
    \end{equation*}
    and we conclude the proof.
\end{proof}

\begin{lemma}\label{lm:swmebp:2}
    For any two balls $B_1(\mathbf{c}_1,r_1)$ and $B_2(\mathbf{c}_2,r_2)$ with $B_2 \subset B_1$,
    it holds that $\mu \cdot B_2 \subset \mu \cdot B_1, \forall\mu \geq 1$.
\end{lemma}
\begin{proof}
    First, we can prove $d(\mathbf{c}_1,\mathbf{c}_2) \leq r_1 - r_2$ by contradiction.
    Then, based on the above result, we can see that, $\forall \mathbf{p}_2 \in \mu \cdot B_2$,
    \begin{equation*}
    d(\mathbf{c}_1,\mathbf{p}_2) \leq d(\mathbf{c}_2,\mathbf{p}_2) + d(\mathbf{c}_1,\mathbf{c}_2)
    \leq \mu r_2 + (r_1 - r_2) \leq \mu r_1
    \end{equation*}
    where the last inequality holds from $r_1 \geq r_2$.
    Thus, $\forall \mathbf{p}_2 \in \mu \cdot B_2$, $\mathbf{p}_2 \in \mu \cdot B_1$.
\end{proof}

According to the above results, we prove the following lemma.
\begin{lemma}\label{lm:smooth}
  For any three point sets $P_1,P_2,P_3$ such that $P_2 \subset P_1$ and $P_3 \not\subset P_1$,
  if $\frac{r^*(P_1)}{r^*(P_2)}=z$, then it holds that $\frac{r^*(P_1 \cup P_3)}{r^*(P_2 \cup P_3)} \leq z + \frac{\sqrt{2}}{2}$.
\end{lemma}
\begin{proof}
  Let $z_1=\frac{r^*(P_3)}{r^*(P_2)}$, $z_2=\frac{d(\mathbf{c}^*(P_2),\mathbf{c}^*(P_3))}{r^*(P_2)}$.
  First, we have
  \begin{equation}\label{eq:r1}
    \begin{split}
      r^*(P_1 \cup P_3) & \leq \frac{1}{2}\big(r^*(P_1) + d(\mathbf{c}^*(P_1),\mathbf{c}^*(P_3)) + r^*(P_3)\big) \\
                        & \leq \frac{1}{2}\big( z + \sqrt{z^2 - 1} + z_1 + z_2 \big) \cdot r^*(P_2)
    \end{split}
  \end{equation}
  Next, we consider three cases separately.
  \begin{itemize}
    \item \textbf{Case 1 ($r^*(P_3) + d(\mathbf{c}^*(P_2),\mathbf{c}^*(P_3)) \leq r^*(P_2)$):}
    In this case, we have $r^*(P_2 \cup P_3) = r^*(P_2)$ and $z_1 + z_2 \leq 1$.
    Thus, we can acquire
    \begin{equation*}
      \frac{r^*(P_1 \cup P_3)}{r^*(P_2 \cup P_3)}
      \leq \frac{1}{2}(z + \sqrt{z^2 - 1} + z_1 + z_2)
      \leq \frac{1}{2}(1 + z + \sqrt{z^2 - 1})
      \leq z + \frac{1}{2}
    \end{equation*}
    \item \textbf{Case 2 ($r^*(P_2) + d(\mathbf{c}^*(P_2),\mathbf{c}^*(P_3)) \leq r^*(P_3)$):}
    In this case, we have $r^*(P_2 \cup P_3) = r^*(P_3)$ and $1 + z_2 \leq z_1$.
    Thus, we can acquire
    \begin{equation*}
      \frac{r^*(P_1 \cup P_3)}{r^*(P_2 \cup P_3)}
      \leq \frac{z + \sqrt{z^2 - 1} + z_1 + z_2}{2z_1}
      = f_1(z,z_1,z_2)
    \end{equation*}
    We can transform it into the following problem:
    \begin{align}\label{eq:smooth:f1}
      \max \quad & \quad f_1(z,z_1,z_2) - z \\
      \text{s.t.} \quad & \quad 1 + z_2 \leq z_1, z \geq 1, z_2 \geq 0 \nonumber
    \end{align}
    Solving Problem~\ref{eq:smooth:f1}, we can get $\frac{r^*(P_1 \cup P_3)}{r^*(P_2 \cup P_3)} \leq z + \frac{1}{2}$.
    \item \textbf{Case 3 (Otherwise):}
    It holds that $z_2 > 0$, $1 + z_2 > z_1$, and $z_1 + z_2 > 1$.
    In addition, we have $d(\mathbf{c}^*(P_2),\mathbf{c}^*(P_3))>0$ because the criteria of either Case 1 or 2
    must be satisfied when $d(\mathbf{c}^*(P_2),\mathbf{c}^*(P_3))=0$.
    As $d(\mathbf{c}^*(P_2 \cup P_3),\mathbf{c}^*(P_2))^2 \leq r^*(P_2 \cup P_3)^2 - r^*(P_2)^2$ and
    $d(\mathbf{c}^*(P_2 \cup P_3),\mathbf{c}^*(P_3))^2 \leq r^*(P_2 \cup P_3)^2 - r^*(P_3)^2$
    according to Lemma~\ref{lm:meb:properties}, we can get
    \begin{equation}\label{eq:smooth:p1}
      \sqrt{r^*(P_2 \cup P_3)^2 - r^*(P_2)^2} + \sqrt{r^*(P_2 \cup P_3)^2 - r^*(P_3)^2}
      \geq d(\mathbf{c}^*(P_2),\mathbf{c}^*(P_3))
    \end{equation}
    Considering Equations~\ref{eq:r1} and~\ref{eq:smooth:p1} collectively, we have
    \begin{equation*}
      \frac{r^*(P_1 \cup P_3)}{r^*(P_2 \cup P_3)}
      \leq \frac{z_2(z + \sqrt{z^2 - 1} + z_1 + z_2)}{\sqrt{\big((1-z_1)^2+z_2^2\big)\big((1+z_1)^2+z_2^2\big)}}
      = f_2(z,z_1,z_2)
    \end{equation*}
    We can transform it into the following problem:
    \begin{align}\label{eq:smooth:f2}
      \max \quad & \quad f_2(z,z_1,z_2) - z \\
      \text{s.t.} \quad & \quad 1 + z_2 > z_1, z_1 + z_2 > 1, z \geq 1, z_1 > 0, z_2 > 0 \nonumber
    \end{align}
    Solving Problem~\ref{eq:smooth:f2}, we can get $\frac{r^*(P_1 \cup P_3)}{r^*(P_2 \cup P_3)} \leq z + \frac{\sqrt{2}}{2}$.
  \end{itemize}
  In all the three cases, we prove Lemma~\ref{lm:smooth}.
\end{proof}

Considering previous results collectively, we can prove the approximation ratio of
SWMEB+ in the following theorem.
\begin{theorem}\label{thm:swmebp:approx}
  For any $\mathbf{p} \in W_t$, it holds that $\mathbf{p} \in (9.66+\varepsilon) \cdot \mathsf{MEB}(S_t)$
  where $\varepsilon=O(\sqrt{\varepsilon_1} + \sqrt{\varepsilon_2})$.
\end{theorem}
\begin{proof}
  We consider four different cases of $S_t$ returned by SWMEB+ at time $t$.
  \begin{itemize}
    \item \textbf{Case 1 ($x_1 \geq t-N+1$):} In this case, we have $t<N$,
    $x_1=1$, and $W_t=P[x_1,t]$. Therefore, it holds that $S_t=S[x_1,t]$ and
    $\mathbf{p} \in (\sqrt{2}+\varepsilon_1) \cdot \mathsf{MEB}(S_t)$ for any $\mathbf{p} \in W_t$
    according to Theorem~\ref{thm:append:only:approx}.
    \item \textbf{Case 2 ($x_1 < t-N+1 \wedge x_2=x_1+1$):} In this case, we have $x_2=t-N+1$ and $W_t=P[x_2,t]$.
    Similarly, $S_t=S[x_2,t]$ and $\mathbf{p} \in (\sqrt{2}+\varepsilon_1) \cdot \mathsf{MEB}(S_t)$ for any $\mathbf{p} \in W_t$.
    \item \textbf{Case 3 ($x_1 < t-N+1 \wedge \frac{r[x_1,t]}{r[x_2,t]} \leq 1+\varepsilon_2$):} In this case,
    we have $W_t \subset P[x_1,t]$ and $W_t \subset B^*[x_1,t]=\mathsf{MEB}(P[x_1,t])$.
    Then, we can acquire $1 \leq \frac{r^*[x_1,t]}{r[x_1,t]},\frac{r^*[x_2,t]}{r[x_2,t]} \leq \sqrt{2}+\varepsilon_1$
    from Theorem~\ref{thm:append:only:approx}. Given $\frac{r^*[x_2,t]}{r[x_2,t]}=\mu$,
    it holds that $\mathbf{p} \in \big( \frac{\sqrt{2}+\varepsilon_3}{\mu} +
    \sqrt{\frac{2+\varepsilon_3}{\mu^2}-1} \big) \cdot B^*[x_2,t], \forall \mathbf{p} \in W_t$,
    where $\varepsilon_3=O(\varepsilon_1+\varepsilon_2)$.
    Additionally, we have $B^*[x_2,t] \subset (\mu + \sqrt{\mu^2 - 1}) \cdot B[x_2,t]$ from Lemma~\ref{lm:swmebp:1}.
    According to Lemma~\ref{lm:swmebp:2}, it holds that $\forall \mathbf{p} \in W_t$,
    \begin{equation}\label{eq:swmebp:case3}\textstyle
      \mathbf{p} \in \Big( \frac{\sqrt{2}+\varepsilon_3}{\mu} +
      \sqrt{\frac{2+\varepsilon_3}{\mu^2}-1} \Big)(\mu + \sqrt{\mu^2 - 1}) \cdot B[x_2,t]
    \end{equation}
    By finding the maximum of
    $f(\mu)=\Big(\sqrt{\frac{2+\varepsilon_3}{\mu^2}-1} \Big)(\mu + \sqrt{\mu^2 - 1})$
    in Equation~\ref{eq:swmebp:case3} on range $[1,\sqrt{2}+\varepsilon_1]$,
    we can acquire $\mathbf{p} \in \big(3.36+\sqrt{\varepsilon_3}\big) \cdot B[x_2,t]$ for any $\mathbf{p} \in W_t$.
    \item \textbf{Case 4 ($x_1 < t-N+1 \wedge \frac{r[x_1,t]}{r[x_2,t]} > 1+\varepsilon_2$):}
    At time $t'<t$ when $x_1$ and $x_2$ become neighboring indices, it holds that
    $\frac{r[x_1,t']}{r[x_2,t']} \leq 1+\varepsilon_2$. Then,
    $\frac{r^*[x_1,t']}{r^*[x_2,t']} \leq (1+\varepsilon_2)(\sqrt{2}+\varepsilon_1)$
    where $r^*[x_1,t']$ and $r^*[x_2,t']$ are the radii of the MEBs for $P[x_1,t']$
    and $P[x_2,t']$ respectively. Furthermore, according to Lemma~\ref{lm:smooth}, we have
    $\frac{r^*[x_1,t]}{r^*[x_2,t]} \leq (1+\varepsilon_2)(\sqrt{2}+\varepsilon_1) + \frac{\sqrt{2}}{2}$.
    According to Lemmas~\ref{lm:swmebp:1} and~\ref{lm:swmebp:2},
    we can acquire that $\forall \mathbf{p} \in W_t$,
    \begin{equation}\label{eq:swmebp:ratio:1}
      \mathbf{p} \in \big(4 + O(\sqrt{\varepsilon_1}+\sqrt{\varepsilon_2}) \big) \cdot B^*[x_2,t]
    \end{equation}
    Moreover, we have $\frac{r^*[x_2,t]}{r[x_2,t]} \leq \sqrt{2}+\varepsilon_1$
    and thus
    \begin{equation}\label{eq:swmebp:ratio:2}
      B^*[x_2,t] \subset \big(\sqrt{2} + 1 + O(\sqrt{\varepsilon_1}) \big) \cdot B[x_2,t]
    \end{equation}
    Combining Equations~\ref{eq:swmebp:ratio:1} and~\ref{eq:swmebp:ratio:2},
    we finally get $\mathbf{p} \in \big(4\sqrt{2} + 4+ \varepsilon \big) \cdot B[x_2,t]
    \approx (9.66 + \varepsilon) \cdot B[x_2,t], \forall\mathbf{p} \in W_t$,
    where $\varepsilon=O(\sqrt{\varepsilon_1}+\sqrt{\varepsilon_2})$.
  \end{itemize}
  Considering all above cases collectively, we conclude the proof.
\end{proof}

According to Theorem~\ref{thm:swmebp:approx}, SWMEB+ can always return a
$(9.66+\varepsilon)$-$\mathsf{Coreset}(W_t)$ at any time $t$.
In practice, since $r[x_1,t]/r[x_2,t]$ is bounded by $1+O(\varepsilon_2)$ in almost all cases,
the approximation ratio of SWMEB+ can be improved to $(3.36+\varepsilon)$ accordingly.
Furthermore, for any $x_i \in X_t$ ($i \leq s-2$), either $r[x_{i+1},t]$ or $r[x_{i+2},t]$ is less than
$(1+\varepsilon_2)r[x_i,t]$.
In addition, it holds that $r[x_1,t] \leq d_{max}$
and $r[x_{s-1},t] \geq 0.5 \cdot d_{min}$.
Therefore, the number of indices in $X_t$ is $O(\frac{\log \theta}{\varepsilon})$.
Accordingly, the time complexity for SWMEB+ to update each point is $O(\frac{m\log^2 \theta}{\varepsilon^4})$
while the number of points stored by SWMEB+ is $O(\frac{\log^2 \theta}{\varepsilon^3})$,
both of which are independent of $N$.

\subsection{Generalization to Kernelized MEB}\label{subsec:kernel:meb}

In real-world
applications~\cite{DBLP:journals/jmlr/TsangKC05,DBLP:journals/tfs/ChungDW09,DBLP:conf/isnn/WeiL08},
it is required to compute the coreset for MEB in a reproducing kernel Hilbert space (RKHS)
instead of Euclidean space.
Given a symmetric positive definite kernel $k(\cdot,\cdot): \mathbb{R}^m \times \mathbb{R}^m \rightarrow \mathbb{R}$
and its associated feature mapping $\varphi(\cdot)$
where $k(\mathbf{p},\mathbf{q})=\langle \varphi(\mathbf{p}),\varphi(\mathbf{q}) \rangle$
for any $\mathbf{p},\mathbf{q} \in \mathbb{R}^m$,
the kernelized MEB of a set of points $P$ is the smallest ball $B^*(\mathbf{c}^*,r^*)$ in the RKHS
such that the maximum distance from $\mathbf{c}^*$ to $\varphi(\mathbf{p})$ is no greater than $r^*$,
which can be formulated as follows:
\begin{equation}\label{eq:kernelized:meb:primal}
  \min_{\mathbf{c}, r}\  r^2 \quad \text{s.t.} \quad
  \big(\mathbf{c}-\varphi(\mathbf{p}_i)\big)'\big(\mathbf{c}-\varphi(\mathbf{p}_i)\big) \leq r^2,
  \forall \mathbf{p}_i \in P
\end{equation}
However, it is impractical to solve Problem~\ref{eq:kernelized:meb:primal}
directly in the primal form due to the infinite dimensionality of RKHS.
We transform Problem~\ref{eq:kernelized:meb:primal} into the dual form
as follows:
\begin{equation}\label{eq:kernelized:meb:dual}
  \max_{\bm{\alpha}}\  \bm{\alpha}' \text{diag}(\mathbf{K}) - \bm{\alpha}'\mathbf{K}\bm{\alpha}
  \quad \text{s.t.} \quad \bm{\alpha} \geq \mathbf{0}, \bm{\alpha}'\mathbf{1}=1
\end{equation}
where $\bm{\alpha}=[\alpha_{1},\ldots,\alpha_{n}]'$ is the $n$-dimensional Lagrange multiplier vector,
$\mathbf{0}=[0,\ldots,0]'$ and $\mathbf{1}=[1,\ldots,1]'$ are both $n$-dimensional vectors,
$\mathbf{K}=[k(\mathbf{p}_i,\mathbf{p}_j)]_{i,j=1}^{n}$ is the $n \times n$ kernel matrix
of $P$, and $\text{diag}(\mathbf{K})$ is the diagonal of $\mathbf{K}$.
Problem~\ref{eq:kernelized:meb:dual} is known to be a quadratic programming~\cite{Boyd:2004:Convex}
(QP) problem. According to the KKT conditions~\cite{Boyd:2004:Convex},
the kernelized MEB $B^*$ of $P$,
can be recovered from the Lagrange multiplier vector $\bm{\alpha}$ as follows:
\begin{equation}\label{eq:kernelized:meb:recovery}
  \mathbf{c}^*=\sum_{i=1}^{n}\alpha_{i} \cdot \varphi(\mathbf{p}_i),
  \  (r^*)^2=\bm{\alpha}'\text{diag}(\mathbf{K})-\bm{\alpha}'\mathbf{K}\bm{\alpha}
\end{equation}
Here the center $\mathbf{c}^*$ is represented implicitly
by each point $\mathbf{p}_i$ in $P$ and the corresponding $\alpha_{i}$'s.
Then, the distance between $\mathbf{c}^*$ and $\varphi(\mathbf{q})$ for any $\mathbf{q} \in \mathbb{R}^m$
can be computed by:
\begin{equation}\label{eq:kernelized:distance}
  d(\mathbf{c}^*,\varphi(\mathbf{q}))^2 = \sum_{i,j=1}^{n} \alpha_{i}\alpha_{j}k(\mathbf{p}_i,\mathbf{p}_j) + k(\mathbf{q},\mathbf{q}) - 2 \sum_{i=1}^{n} \alpha_{i}k(\mathbf{p}_{i},\mathbf{q})
\end{equation}

Next, we introduce how to generalize AOMEB in Algorithm~\ref{alg:coreset:append:only} to maintain coresets for kernelized MEB.
First, to represent the center $\mathbf{c}_t$ and compute the distance from $\varphi(\mathbf{p}_t)$ to the center,
it always keeps the Lagrange multiplier vector $\bm{\alpha}$.
In Line~\ref{line:append:only:init}, $\alpha_1$ is initialized to $1$ so that $\mathbf{c}_1=\varphi(\mathbf{p}_1)$.
Then, in Line~\ref{line:append:only:streaming:s}, Equation~\ref{eq:kernelized:distance} is used
to compute the distance between $\mathbf{c}_{t-1}$ and $\varphi(\mathbf{p}_t)$.
If $\mathbf{p}_t$ is added to $S_t$, it will re-optimize Problem~\ref{eq:kernelized:meb:dual} on $S_t$
to adjust $\bm{\alpha}$ so that $B_t=\mathsf{MEB}(S_t)$ in Line~\ref{line:offline:iter:t}.
Specifically, the Frank-Wolfe algorithm~\cite{DBLP:journals/siamjo/Yildirim08,DBLP:journals/talg/Clarkson10}
is used to solve Problem~\ref{eq:kernelized:meb:dual}
as it is efficient for QP problems with unit simplex constraints.
SWMEB and SWMEB+ can also maintain coresets for kernelized MEB
by using the generalized AOMEB instance in Algorithms~\ref{alg:swmeb} and~\ref{alg:swmeb:plus}.

Theoretically, the generalized algorithms have the same approximation ratios and coreset sizes as the original ones.
However, the time complexity will increase by a factor of $\frac{1}{\varepsilon}$
because the time to compute the distance from $\mathbf{c}_{t-1}$ to $\varphi(\mathbf{p}_t)$ using Equation~\ref{eq:kernelized:distance}
is $O(\frac{m}{\varepsilon})$ instead of $O(m)$.

\subsection{Discussion}

For ease of presentation, we describe Algorithms~\ref{alg:coreset:append:only}--\ref{alg:swmeb:plus}
in the single-update-mode where the coreset is maintained for every new point.
In practice, it is not required to update the coreset at such an intense rate.
Here we discuss how to adapt these algorithms for the \emph{mini-batch-mode}.

In the mini-batch-mode, given a batch size $b$, each update will add $b$ new points
while deleting the earliest $b$ points at the same time.
The adaptations of AOMEB in Algorithm~\ref{alg:coreset:append:only}
for the mini-batch-mode are as follows:
(1) In Line~\ref{line:append:only:init}, an initial coreset $S_b$ is built
for the first $b$ points using CoreMEB in Algorithm~\ref{alg:coreset:offline}.
(2) In Lines~\ref{line:append:only:streaming:s}--\ref{line:append:only:streaming:t},
the coreset is updated for a batch of $b$ points collectively.
Specifically, it adds the points not contained in $(1+\varepsilon_1) \cdot B_{t-b}$
in the batch to $S_t$ and then updates $B_t$.
To adapt SWMEB and SWMEB+ for the mini-batch-mode, each AOMEB instance
should run in the mini-batch-mode as shown above.
In addition, the indices and AOMEB instances in SWMEB and SWMEB+
are created and updated for batches instead of points.
Note that the approximation ratio and coreset size will remain the same in the mini-batch-mode
but the efficiency will be improved as fewer indices are created.

\section{Experiments}\label{sec:experiments}

In this section we evaluate the empirical performance of our proposed algorithms
on real-world and synthetic datasets.
First of all, we will introduce the experimental setup in Section~\ref{subsec:exp:setup}.
Then, we will present the experimental results on effectiveness and efficiency
in Section~\ref{subsec:exp:result:euclidean}.
Finally, the experimental results on scalability
are presented in Section~\ref{subsec:exp:result:kernel}.

\subsection{Experimental Setup}\label{subsec:exp:setup}

\textbf{Algorithms.}
We compare the following eight algorithms for computing MEBs
or building coresets for MEB in our experiments.
\begin{itemize}
    \item \textbf{COMEB}~\cite{DBLP:conf/esa/FischerGK03}: a combinatorial algorithm for computing
    the \emph{exact} MEB of a set of points in the Euclidean space. It has an exponential time complexity
    w.r.t. the dimension $m$. Moreover, it is not applicable to kernelized MEB.
    \item \textbf{CoreMEB}~\cite{DBLP:journals/comgeo/BadoiuC08}: a batch algorithm for constructing
    a $(1+\varepsilon)$-approximate coreset for the MEB of a set of points. The procedure
    is as described in Algorithm~\ref{alg:coreset:offline}.
    \item \textbf{SSMEB}~\cite{DBLP:conf/cccg/Zarrabi-ZadehC06}: a $1.5$-approximation algorithm for computing a MEB
    in an append-only stream. We adopt the method described in~\cite{DBLP:conf/ijcai/RaiDV09}
    to compute a kernelized MEB by SSMEB.
    \item \textbf{Blurred Ball Cover (BBC)}~\cite{DBLP:conf/soda/AgarwalS10}: an append-only streaming
    algorithm to maintain a $(\sqrt{2}+\varepsilon)$-approximate coreset for MEB.
    \item \textbf{DyMEB}~\cite{DBLP:journals/comgeo/ChanP14}: a $1.22$-approximate dynamic algorithm
    for MEB computation. It keeps a data structure that
    permits to insert/delete random points without fully reconstructions.
    \item \textbf{AOMEB}: our append-only streaming algorithm presented in Section~\ref{subsec:append:only}.
    It has the same $(\sqrt{2}+\varepsilon)$-approximation ratio as BBC.
    \item \textbf{SWMEB}: our first sliding-window algorithm presented in Section~\ref{subsec:swmeb}.
    It can maintain a $(\sqrt{2}+\varepsilon)$-coreset for MEB over the sliding window.
    \item \textbf{SWMEB+}: our second sliding-window algorithm presented in Section~\ref{subsec:swmeb:plus}.
    It has higher efficiency than SWMEB at the expense of a worse approximation ratio.
\end{itemize}
We do not compare with the algorithms in~\cite{DBLP:journals/jacm/AgarwalHV04,DBLP:journals/comgeo/Chan06}
because they cannot scale to the datasets with $m>10$.
In our experiments, all algorithms run in the mini-batch-mode with batch size $b=100$.
Furthermore, all algorithms, except for SWMEB and SWMEB+, cannot directly work in the sliding window model.
The batch and append-only streaming algorithms store the entire sliding window
and rerun from scratch for each update.
DyMEB also stores the entire sliding window for tracking the expired point to delete.
For each update, it must execute one deletion/insertion for every expired/arrival point in a mini-batch
to maintain the coreset for MEB w.r.t.~the up-to-date window.
The implementations of our algorithms are available
at \url{https://github.com/yhwang1990/SW-MEB}.

\begin{table}[t]
    \centering
    \small
    \caption{Statistics of datasets}
    \label{tbl:dataset}
    \begin{tabular}{|c|c|c|c|c|c|}
        \hline
        \textbf{dataset}   & \textbf{source} & \textbf{size} & $m$ & $\theta$ & $\gamma$ \\
        \hline
        \textsf{Census}    & UCI             & 2,458,285  & 68  & 228.5 & 8984.58  \\
        \textsf{CovType}   & LIBSVM          & 297,711    & 54  & 246.8 & 3.734    \\
        \textsf{GIST}      & TEXMAX          & 1,000,000  & 960 & 45.77  & 4.0409   \\
        \textsf{Gowalla}   & SNAP            & 6,442,892  & 2   & $\approx$3000 & 7455.33  \\
        \textsf{HIGGS}     & UCI             & 5,829,123  & 28  & 13.69 & 38.8496  \\
        \textsf{SIFT}      & TEXMAX          & 1,000,000  & 128 & 12.14  & 298919.5 \\
        \textsf{Synthetic} & -               & 10,000,000 & 50  & 2.96  & 100.6    \\
        \hline
    \end{tabular}
\end{table}

\textbf{Datasets.}
The dataset statistics are listed in Table~\ref{tbl:dataset}.
We use 6 real-world datasets and 1 synthetic dataset for evaluation.
All real-world datasets are downloaded from publicly available sources,
e.g., UCI Machine Learning Repository\footnote{\url{https://archive.ics.uci.edu/ml/index.php}},
LIBSVM\footnote{\url{https://www.csie.ntu.edu.tw/~cjlin/libsvmtools/datasets}},
SNAP\footnote{\url{http://snap.stanford.edu}},
and TEXMAX\footnote{\url{http://corpus-texmex.irisa.fr}}.
The generation procedure of the \textsf{Synthetic} dataset is as follows.
We first decide the dimension $m$. By default, we set $m=50$.
For testing the scalability of different algorithms w.r.t.~$m$,
we vary $m$ from 10 to 100 and from 1,000 to 10,000.
Then, we generate a point by drawing the values of $m$ dimensions
from a normal distribution $\mathcal{N}(0,1)$ independently.
For kernelized MEB, we adopt the Gaussian kernel
$k(\mathbf{p}_i,\mathbf{p}_j)=\exp(-d(\mathbf{p}_i,\mathbf{p}_j)^2/\gamma)$
where $\gamma=\frac{1}{n^2}\sum_{i,j=1}^{n} d(\mathbf{p}_i,\mathbf{p}_j)^2$ is the kernel width.
In practice, we sample 10,000 points randomly from each dataset to compute $\gamma$.
The results are also listed in Table~\ref{tbl:dataset}.
Note that the values of $\gamma$ vary with $m$ ($\gamma \approx 2m$) on the \textsf{Synthetic} dataset.
More details on datasets are provided in Appendix~\ref{appendix:datasets}.

In an experiment, all points in a dataset are processed sequentially by each algorithm
in the same order as a stream and the results are recorded for every batch of points.

\textbf{Parameter Setting.}
The default window size $N$ is $10^5$ in all experiments
except the ones for testing the scalability w.r.t.~$N$,
where we vary $N$ from $10^5$ to $10^6$.
The parameter $\varepsilon_1$ in AOMEB, SWMEB, and SWMEB+
(as well as $\varepsilon$ in CoreMEB, BBC, and DyMEB) is
$10^{-3}$ for Euclidean MEB and $10^{-4}$ for kernelized MEB.
Then, we use $\varepsilon_2=0.1$ in SWMEB and
$\varepsilon_2=\min(4^{i-1} \cdot \frac{\varepsilon_1}{10}, 0.1)$ for each $x_i \in X_t$ in SWMEB+.
Finally, the partition size $L$ in SWMEB is $\frac{N}{10}$.
The procedure of parameter tuning is shown in Appendix~\ref{appendix:parameter}.

\textbf{Environment.}
All experiments are conducted on a server running Ubuntu 16.04
with a 1.9GHz processor and 128 GB memory.
The detailed hardware and software configurations are in Appendices~\ref{appendix:hardware} and~\ref{appendix:software}.

\begin{table*}[t]
    \small
    \centering
    \caption{The average errors of different algorithms for Euclidean MEB}
    \label{tbl:error:euclidean}
    \begin{tabular}{c|ccccccc}
        \hline
        \multirow{2}{*}{\textbf{Algorithm}} & \multicolumn{7}{c}{\textbf{average error}}
        \\ \cline{2-8}
        & \textsf{Census} & \textsf{CovType} & \textsf{GIST} & \textsf{Gowalla} & \textsf{HIGGS} & \textsf{SIFT} & \textsf{Synthetic}
        \\ \hline
        \textbf{COMEB}   & 0        & 0        & 0        & 0        & 0        & 0        & 0        \\
        \textbf{CoreMEB} & 6.03e-04 & 3.66e-04 & 4.32e-04 & 2.01e-04 & 4.59e-04 & 4.97e-04 & 4.49e-04 \\
        \textbf{SSMEB}   & 5.90e-02 & 1.56e-01 & 1.10e-01 & 7.19e-03 & 1.59e-01 & 1.61e-01 & 9.49e-02 \\
        \textbf{BBC}     & 1.59e-03 & 8.05e-03 & 4.47e-04 & 1.53e-03 & 1.33e-02 & 5.83e-03 & 2.05e-02 \\
        \textbf{DyMEB}   & 9.14e-04 & 3.27e-03 & 4.51e-04 & 1.98e-04 & 2.19e-03 & 4.34e-03 & 5.96e-03 \\
        \hline
        \textbf{AOMEB}   & 9.55e-04 & 4.55e-03 & 4.55e-04 & 2.23e-04 & 4.73e-03 & 2.86e-03 & 1.14e-02 \\
        \textbf{SWMEB}   & 6.12e-04 & 8.36e-03 & 2.31e-04 & 2.57e-04 & 3.87e-03 & 2.94e-03 & 1.11e-02 \\
        \textbf{SWMEB+}  & 8.98e-04 & 4.17e-03 & 1.81e-03 & 1.93e-04 & 9.67e-03 & 3.01e-03 & 1.52e-02 \\
        \hline
    \end{tabular}
\end{table*}

\begin{table*}[t]
    \small
    \centering
    \caption{The average update time of different algorithms for Euclidean MEB}
    \label{tbl:time:euclidean}
    \begin{tabular}{c|ccccccc}
        \hline
        \multirow{2}{*}{\textbf{Algorithm}} & \multicolumn{7}{c}{\textbf{average update time (ms)}}
        \\ \cline{2-8}
        & \textsf{Census} & \textsf{CovType} & \textsf{GIST} & \textsf{Gowalla} & \textsf{HIGGS} & \textsf{SIFT} & \textsf{Synthetic}
        \\ \hline
        \textbf{COMEB}   & 1150.7   & 1914.8   & 3813.7   & 13.52    & 146.67   & 12861.8  & 2793.7   \\
        \textbf{CoreMEB} & 643.1    & 1275.1   & 2647.2   & 30.19    & 337.72   & 2789.1   & 1698.8   \\
        \textbf{SSMEB}   & 43.75    & 44.11    & 262.8    & 24.55    & 45.66    & 74.16    & 54.46    \\
        \textbf{BBC}     & 193.4    & 767.4    & 961.8    & 37.61    & 118.19   & 6807.2   & 1150.4   \\
        \textbf{DyMEB}   & 1639.4   & 2725.9   & 7498.8   & 80.32    & 3483.2   & 9195.1   & 2786.3   \\
        \hline
        \textbf{AOMEB}   & 95.74    & 507.3    & 1755.5   & 17.82    & 125.48   & 1413.9   & 480.63   \\
        \textbf{SWMEB}   & 2.129    & 25.08    & 127.77   & 0.1593   & 4.861    & 57.61    & 19.9     \\
        \textbf{SWMEB+}  & 1.467    & 5.414    & 72.31    & 0.1887   & 2.592    & 14.37    & 5.679    \\
        \hline
    \end{tabular}
\end{table*}

\begin{table*}[t]
    \small
    \centering
    \caption{The average errors of different algorithms for kernelized MEB}
    \label{tbl:error:kernel}
    \begin{tabular}{c|ccccccc}
        \hline
        \multirow{2}{*}{\textbf{Algorithm}} & \multicolumn{7}{c}{\textbf{average error}}
        \\ \cline{2-8}
        & \textsf{Census} & \textsf{CovType} & \textsf{GIST} & \textsf{Gowalla} & \textsf{HIGGS} & \textsf{SIFT} & \textsf{Synthetic}
        \\ \hline
        \textbf{CoreMEB} & 9.31e-05 & 9.16e-05 & 9.76e-05 & 7.97e-05 & 9.64e-05 & 9.40e-05 & 9.53e-05 \\
        \textbf{SSMEB}   & 2.22e-01 & 1.73e-01 & 1.83e-01 & 2.26e-01 & 2.08e-01 & 1.67e-01 & 1.40e-01 \\
        \textbf{BBC}     & 1.21e-03 & 1.55e-03 & 1.78e-05 & 1.15e-02 & 9.02e-04 & 1.60e-03 & 1.48e-03 \\
        \textbf{DyMEB}   & 1.62e-04 & 6.07e-05 & 1.18e-04 & 2.07e-03 & 1.57e-04 & 9.91e-05 & 4.26e-05 \\
        \hline
        \textbf{AOMEB}   & 5.63e-04 & 6.02e-04 & 1.62e-06 & 1.02e-04 & 6.08e-04 & 7.55e-04 & 4.59e-04 \\
        \textbf{SWMEB}   & 7.77e-04 & 4.71e-03 & 5.01e-03 & 1.19e-04 & 5.82e-03 & 3.12e-03 & 4.99e-03 \\
        \textbf{SWMEB+}  & 5.86e-04 & 1.01e-03 & 4.85e-04 & 1.67e-04 & 2.10e-03 & 6.78e-04 & 1.61e-03 \\
        \hline
    \end{tabular}
\end{table*}

\begin{table*}[t]
    \small
    \centering
    \caption{The average update time of different algorithms for kernelized MEB}
    \label{tbl:time:kernel}
    \begin{tabular}{c|ccccccc}
        \hline
        \multirow{2}{*}{\textbf{Algorithm}} & \multicolumn{7}{c}{\textbf{average update time (ms)}}
        \\ \cline{2-8}
        & \textsf{Census} & \textsf{CovType} & \textsf{GIST} & \textsf{Gowalla} & \textsf{HIGGS} & \textsf{SIFT} & \textsf{Synthetic}
        \\ \hline
        \textbf{CoreMEB} & 75041.6  & 91571.8  & 36522    & 5813.1   & 87875.1  & 149386   & 76223.1  \\
        \textbf{SSMEB}   & 208.67   & 251.12   & 2893.9   & 118.71   & 169.92   & 395.5    & 255.25   \\
        \textbf{BBC}     & 14914.5  & 34728.1  & 704506   & 410.30   & 59789.6  & 148539   & 44383.9  \\
        \textbf{DyMEB}   & 106093   & 256616   & 2484757  & 7142.1   & 345400   & 1333728  & 513587   \\
        \hline
        \textbf{AOMEB}   & 2002.1   & 5716.6   & 42961.2  & 253.87   & 6471.1   & 19163.7  & 10556.2  \\
        \textbf{SWMEB}   & 146.09   & 545.9    & 4550.3   & 18.118   & 606.2    & 1705.7   & 1077.2   \\
        \textbf{SWMEB+}  & 26.847   & 52.06    & 372.4    & 4.769    & 55.85    & 130.46   & 87.21    \\
        \hline
    \end{tabular}
\end{table*}

\subsection{Effectiveness and Efficiency}\label{subsec:exp:result:euclidean}

\begin{figure*}[t]
  \centering
  \begin{subfigure}{0.96\textwidth}
    \includegraphics[width=\textwidth]{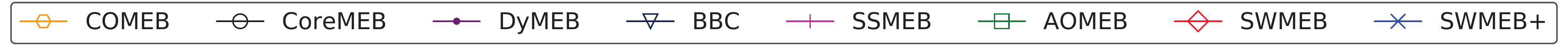}
  \end{subfigure}
  \vskip\baselineskip
  \vspace{-1em}
  \begin{subfigure}{0.4\textwidth}
    \includegraphics[width=\textwidth]{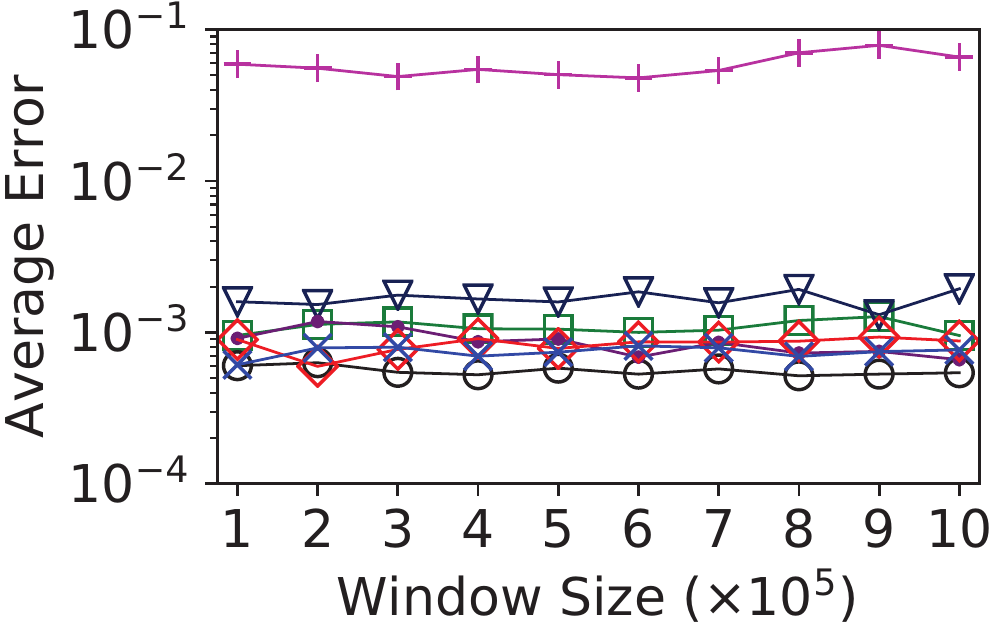}
  \end{subfigure}
  \begin{subfigure}{0.4\textwidth}
    \includegraphics[width=\textwidth]{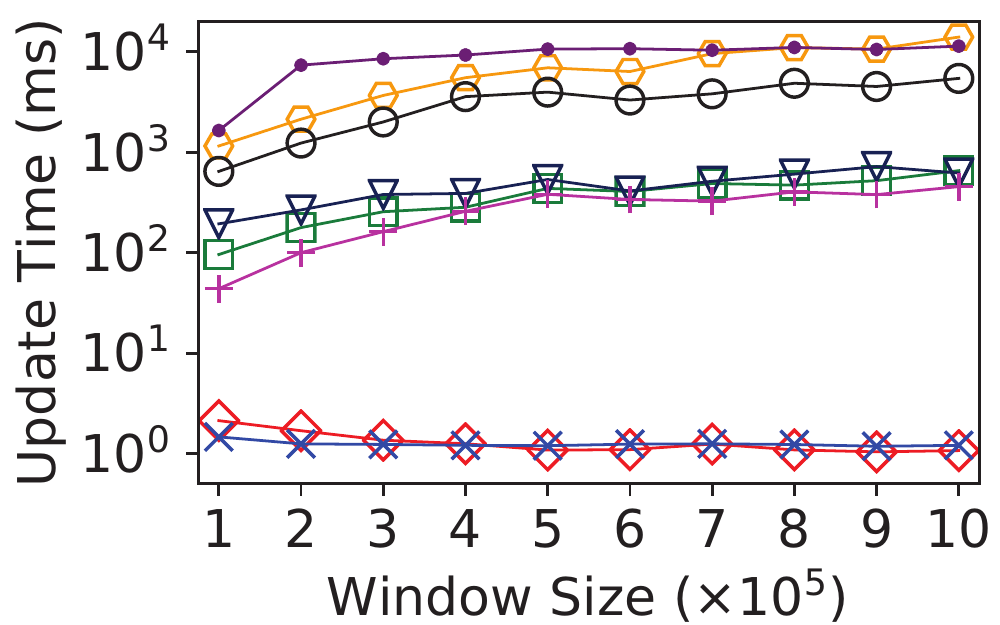}
  \end{subfigure}
  \vskip\baselineskip
  \vspace{-1em}
  \begin{subfigure}{0.4\textwidth}
    \includegraphics[width=\textwidth]{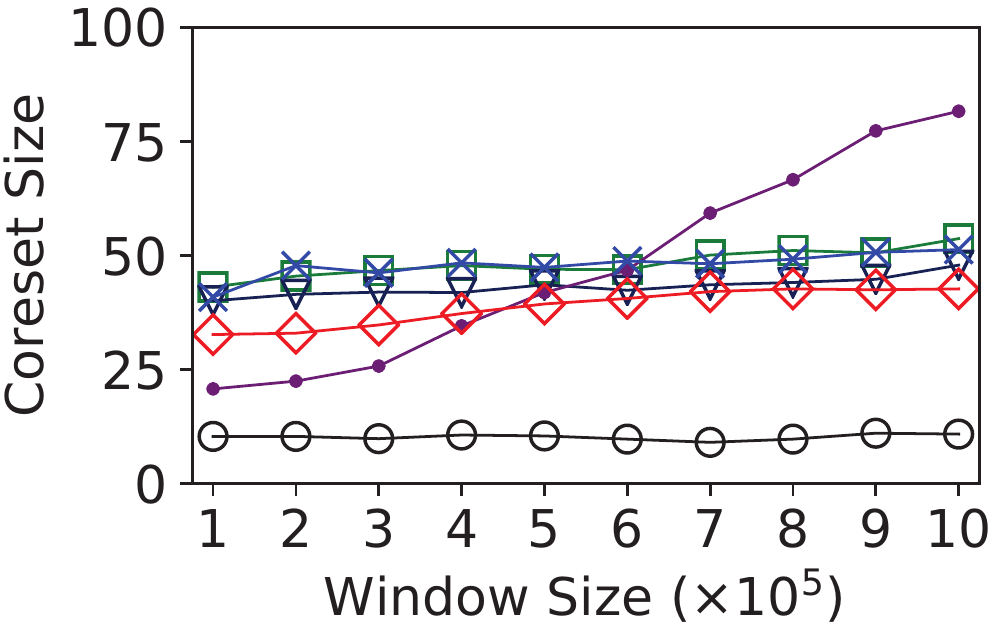}
  \end{subfigure}
  \begin{subfigure}{0.4\textwidth}
    \includegraphics[width=\textwidth]{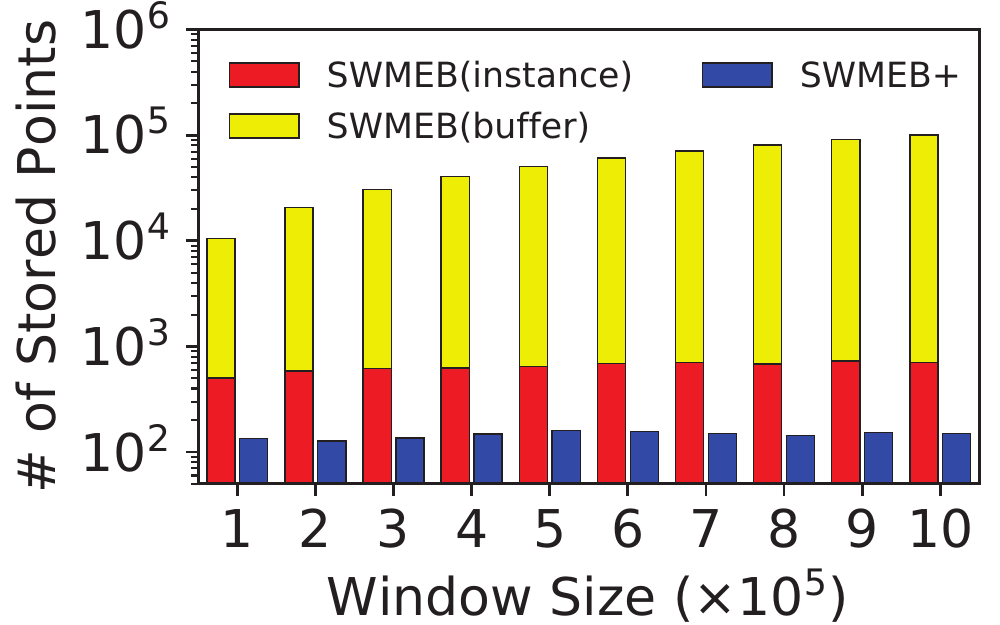}
  \end{subfigure}
  \caption{The performance for Euclidean MEB with varying $N$ on the Census dataset.}
  \label{fig:Euclidean:N}
\end{figure*}

\begin{figure*}[t]
  \centering
  \begin{subfigure}{0.96\textwidth}
      \includegraphics[width=\textwidth]{legend-eps-converted-to.pdf}
  \end{subfigure}
  \vskip\baselineskip
  \vspace{-1em}
  \begin{subfigure}{0.4\textwidth}
    \includegraphics[width=\textwidth]{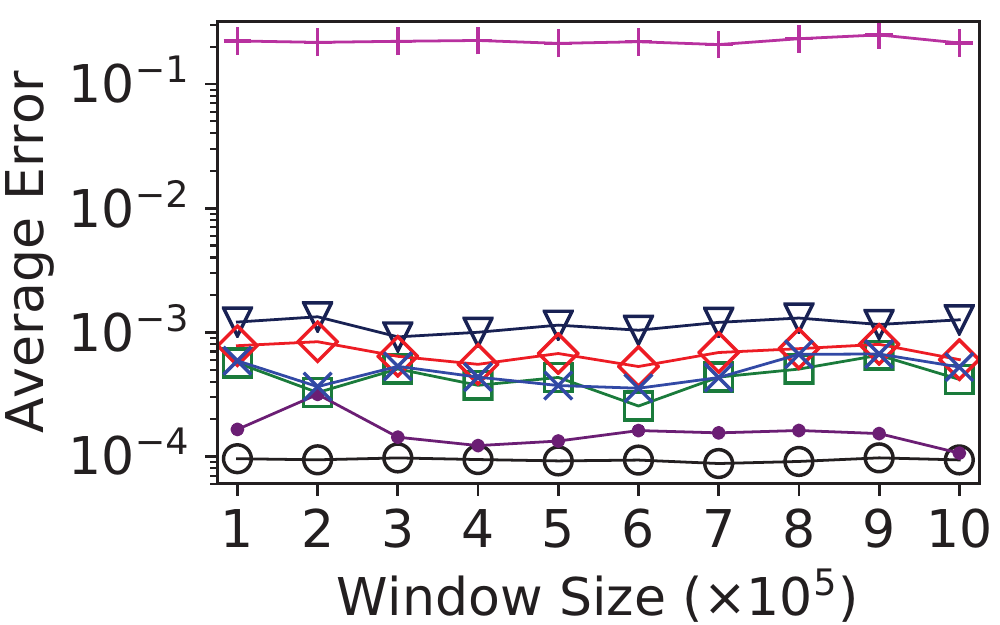}
  \end{subfigure}
  \begin{subfigure}{0.4\textwidth}
    \includegraphics[width=\textwidth]{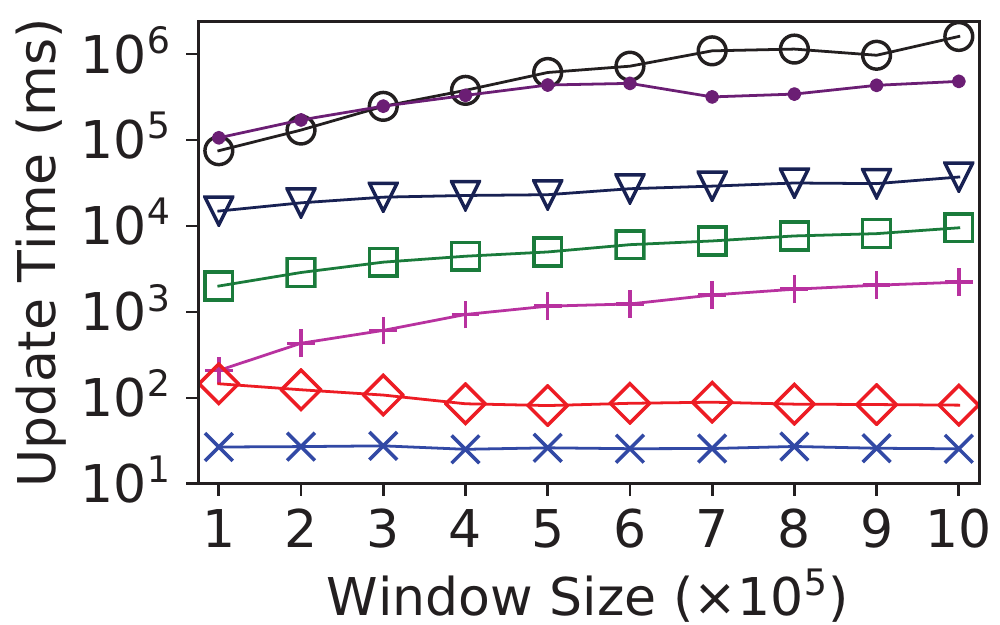}
  \end{subfigure}
  \vskip\baselineskip
  \vspace{-1em}
  \begin{subfigure}{0.4\textwidth}
    \includegraphics[width=\textwidth]{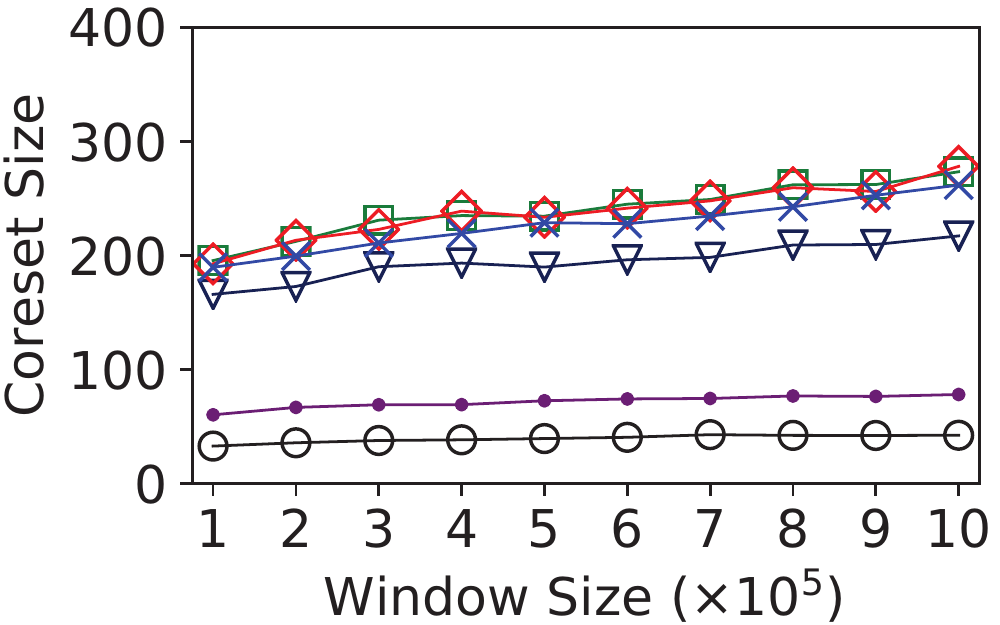}
  \end{subfigure}
  \begin{subfigure}{0.4\textwidth}
    \includegraphics[width=\textwidth]{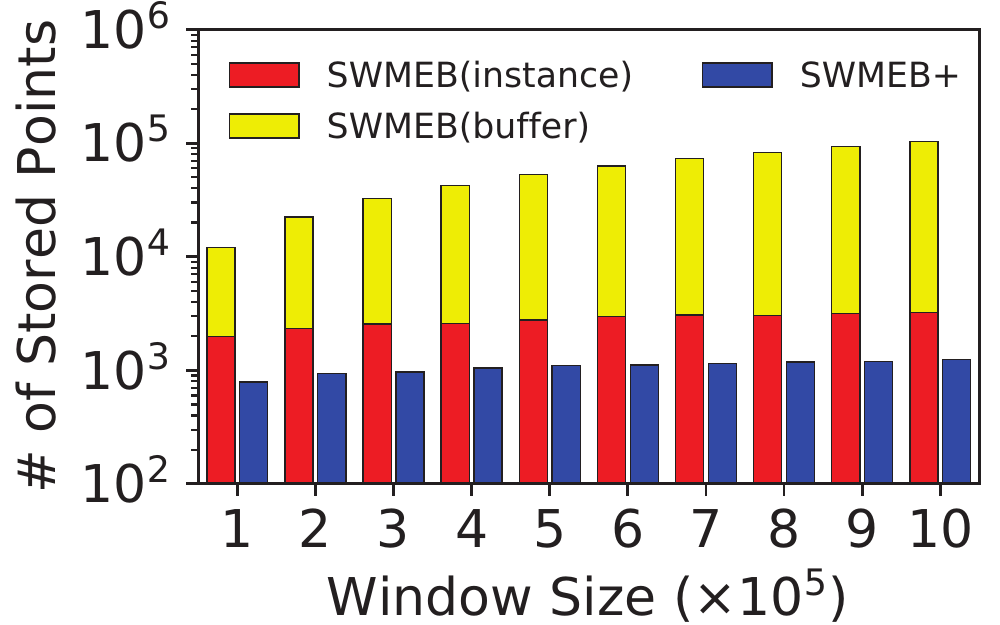}
  \end{subfigure}
  \caption{The performance for kernelized MEB with varying $N$ on the Census dataset.}
  \label{fig:kernel:N}
\end{figure*}

\begin{figure*}[t]
  \centering
  \begin{subfigure}{0.96\textwidth}
      \includegraphics[width=\textwidth]{legend-eps-converted-to.pdf}
  \end{subfigure}
  \vskip\baselineskip
  \vspace{-1em}
  \begin{subfigure}{0.4\textwidth}
    \includegraphics[width=\textwidth]{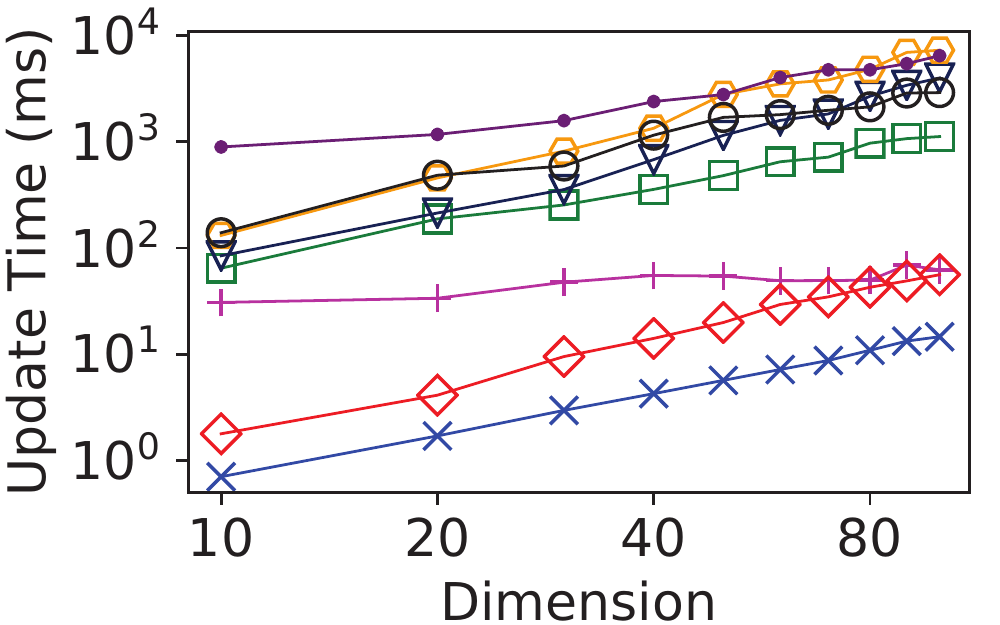}
  \end{subfigure}
  \begin{subfigure}{0.4\textwidth}
    \includegraphics[width=\textwidth]{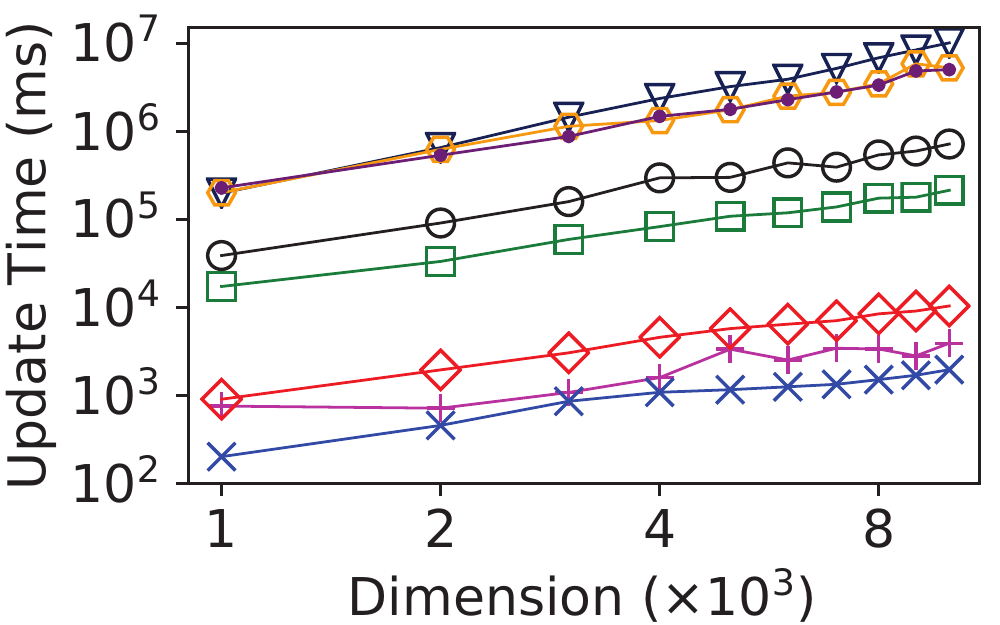}
  \end{subfigure}
  \vskip\baselineskip
  \vspace{-1em}
  \begin{subfigure}{0.4\textwidth}
    \includegraphics[width=\textwidth]{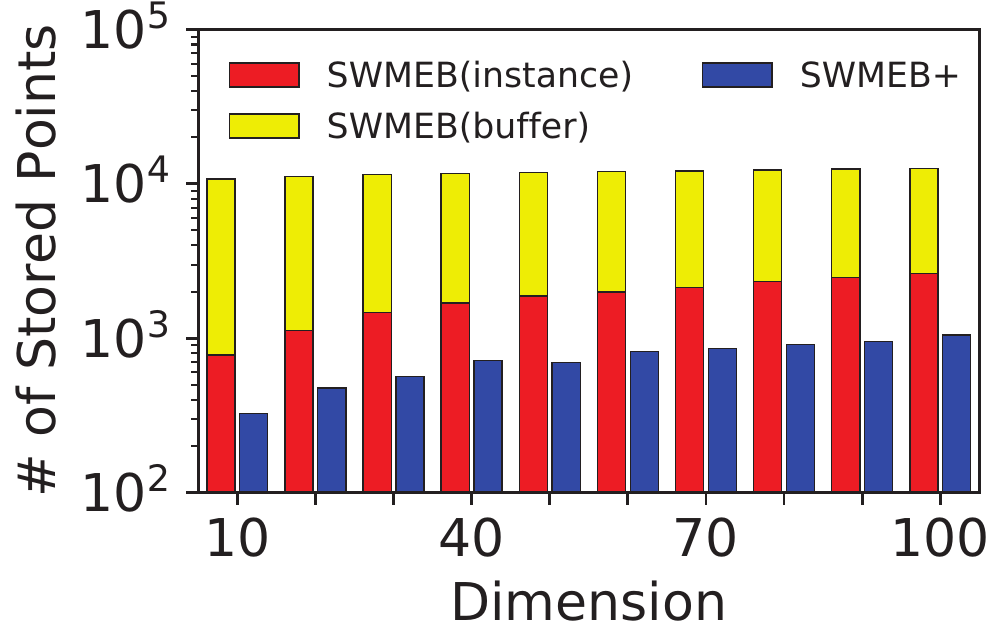}
  \end{subfigure}
  \begin{subfigure}{0.4\textwidth}
    \includegraphics[width=\textwidth]{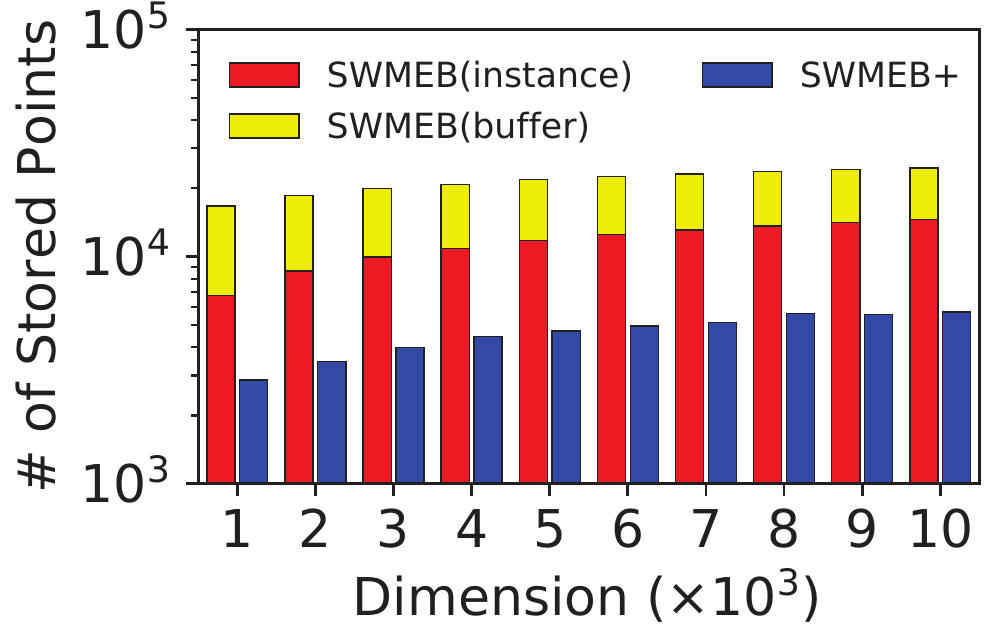}
  \end{subfigure}
  \caption{The performance for Euclidean MEB with varying $m$ on the Synthetic dataset.}
  \label{fig:Euclidean:m}
\end{figure*}

The result quality of different algorithms is evaluated by \emph{average error}
computed as follows. For a sliding window $W_t$, each algorithm
(except COMEB and SSMEB) returns a coreset $S_t$ for $\mathsf{MEB}(W_t)$.
The errors of coreset-based algorithms are acquired based on the definition of \emph{coresets}:
we compute the minimal $\lambda'$ such that $W_t \subset \lambda' \cdot \mathsf{MEB}(S_t)$,
and use $\varepsilon'=\frac{\lambda' \cdot r^*(S_t) - r^*(W_t)}{r^*(W_t)}$ for the relative error.
Since SSMEB directly returns MEBs,
we compute its relative error according to the definition of approximate MEB, i.e.,
$\varepsilon'=\frac{r(B'_t)-r^*(W_t)}{r^*(W_t)}$ where $B'_t$ is the approximate MEB for $W_t$.
It is noted that for kernelized MEB we use the radius of the MEB returned by CoreMEB
when $\varepsilon=10^{-9}$ (i.e., the relative error is within $10^{-9}$)
as $r^*(W_t)$ since exact MEBs are intractable in a RKHS~\cite{DBLP:journals/jmlr/TsangKC05}.
On each dataset, we take $100$ timestamps over the stream,
obtain the result of each algorithm for the sliding window at each sampled timestamp,
compute the relative errors of obtained results,
and use the \emph{average error} as the quality metric.
The efficiency of different algorithms is evaluated by \emph{average update time}.
We record the CPU time of each algorithm to update every batch of $100$ points and
compute the average time per update over the entire dataset.

The \emph{average errors} and \emph{average update time} of different algorithms
for Euclidean and kernelized MEBs are presented
in Tables~\ref{tbl:error:euclidean}--\ref{tbl:time:kernel} respectively.
Note that COMEB cannot be applied to kernelized MEB and thus is not included
in Tables~\ref{tbl:error:kernel} and~\ref{tbl:time:kernel}.

In general, all algorithms except SSMEB can provide MEBs or coresets for MEB
with rather small errors (at most $1.5\%$) in all cases.
Furthermore, the update time for kernelized MEBs is
one to two orders of magnitude longer than that for Euclidean MEB
due to (1) higher time complexity for distance evaluation,
(2) larger coresets caused by the infinite dimensionality of RKHS,
and (3) smaller $\varepsilon_1$ (or $\varepsilon$) used.
In terms of effectiveness, the error of COMEB for Euclidean MEB is always $0$ because it can return exact results.
Additionally, as CoreMEB guarantees a $(1+\varepsilon)$-approximation ratio theoretically, the errors of the coresets
returned by CoreMEB are less than $10^{-3}$ for Euclidean MEB and $10^{-4}$ for kernelized MEB.
With regard to the efficiency for Euclidean MEB, COMEB only runs faster than CoreMEB on low-dimensional datasets
(i.e., \textsf{Gowalla} and \textsf{HIGGS}).
CoreMEB can always outperform COMEB when $m \geq 30$.
The reason is that COMEB has a higher time complexity than CoreMEB for its exponential dependency on $m$.

There are three append-only streaming algorithms in our experiments, namely BBC, SSMEB, and our proposed AOMEB.
Among them, the result quality of SSMEB is not competitive with any other algorithms, though its update time
is shorter than BBC and AOMEB. This is because the simple geometric method for update in SSMEB
is efficient but largely inaccurate.
The errors of BBC and AOMEB are slightly higher than those of CoreMEB
while they are more efficient than CoreMEB.
These experimental results are consistent with our theoretical analysis:
AOMEB and BBC have a lower $(\sqrt{2}+\varepsilon)$-approximation ratio but
only require a single-pass scan over the window.
Finally, AOMEB can run up to $9$ times faster than BBC while having similar or better coreset quality
because AOMEB maintains fewer MEBs than BBC, which leads to a more efficient update procedure.

The dynamic algorithm, i.e., DyMEB, shows slightly better coreset quality than AOMEB and BBC
but runs even slower than CoreMEB.
There are two reasons for such observations:
first, the data structure maintained by DyMEB
contains all points in $W_t$ for coreset construction, which naturally leads to good quality;
second, the performance of DyMEB depends on the assumption that the probability of deleting
any existing point is equal. However, the sliding window model always deletes the earliest point,
which obviously violates this assumption. As a result, when expired points are deleted,
DyMEB frequently calls for (partial) coreset reconstructions.
In practice, the average update time of DyMEB even exceeds the time to build the coreset from scratch.

Finally, our sliding-window algorithms, namely SWMEB and SWMEB+,
achieve two to four orders of magnitude speedups over CoreMEB across all datasets for both Euclidean and kernelized MEBs.
In addition, they run $10$ to $150$ times faster than AOMEB.
The reason for their superior efficiency
is that they can maintain the coreset incrementally over the sliding window
without rebuilding from scratch.
In terms of effectiveness, the errors of SWMEB and SWMEB+
are slightly higher than those of AOMEB.
Although both algorithms use AOMEB instances to provide the coresets,
the index schemes inevitably cause quality losses since the points between
the beginning of $W_t$ (i.e., $t-N+1$) and the first non-expired index
(i.e., $x_{1,{s_1}}$ in SWMEB and $x_2$ in SWMEB+) are missing from the coresets.
Lastly, SWMEB+ runs up to $14$ times faster than SWMEB
owing to the fact that SWMEB+ maintains fewer indices than SWMEB.
At the same time, SWMEB+ can return coresets with nearly equal quality
to SWMEB, which means that the radius-based index scheme of SWMEB+,
though having a worse theoretical guarantee, is competitive with
the partition-based scheme of SWMEB empirically.

\subsection{Scalability}\label{subsec:exp:result:kernel}

We compare the scalability of different algorithms with varying the window size $N$ and dimension $m$.
The performance for Euclidean and kernelized MEBs with varying $N$ is
shown in Figures~\ref{fig:Euclidean:N} and~\ref{fig:kernel:N}.
Here we only present the results on the \textsf{Census} dataset.
The experimental results on other datasets are shown in Appendix~\ref{appendix:results}.

First, we observe that the average errors of different algorithms basically remain stable w.r.t.~$N$
(around $10^{-3}$ for SWMEB \& SWMEB+).
The update time of all algorithms, except SWMEB and SWMEB+, increases with $N$
because the number of points processed per update is equal to $N$.
For SWMEB and SWMEB+, the number of indices in $X_t$ hardly changes $N$.
Besides, the update frequency of any AOMEB instance decreases over time, when the
coreset grows larger and fewer new points can be added.
Since both algorithms contain ``older'' instances when $N$ is larger,
they take less time for each update on average.
In addition, the coreset sizes of our algorithms are $3$--$5$ times larger than that of CoreMEB
because the \emph{greedy} strategy used by AOMEB inevitably adds some redundant points to coresets.
Despite this, the coreset sizes are at most $0.3\% \cdot N$.
Moreover, the coreset size for kernelized MEBs is around $5$ times
larger then that for Euclidean MEBs due to the infinite dimensionality of RKHS.
In terms of space, SWMEB stores up to $100$ times more points than SWMEB+
since it not only maintains more AOMEB instances that SWMEB+ but also
needs to keep a buffer $Q$, whose size is at most $10\% \cdot N$.
Specifically, SWMEB stores at most $13\% \cdot N$ points while
SWMEB+ only keeps up to $2,000$ points, which barely changes with $N$.

The update time and space usage of different algorithms with varying the dimension $m$ is shown in Figure~\ref{fig:Euclidean:m}.
Here we only present the results for Euclidean MEBs as the trend is generally similar for kernelized MEBs.
As plotted in log-log scale, we can observe that the update time of all algorithms except SSMEB increases almost linearly with $m$.
Nevertheless, SWMEB and SWMEB+ still demonstrate their superiority in efficiency on high-dimensional datasets.
Furthermore, both algorithms store more points when $m$ is larger because the coreset size grows with $m$.
But even when $m=10,000$ the ratios of stored points are at most $25\%$ and $5\%$ for SWMEB and SWMEB+ respectively,
whereas any other algorithms require to store the entire window of points. 

\section{Conclusion}\label{sec:conclusion}

We studied the problem of maintaining a coreset for the MEB of a sliding window of points in this paper.
Firstly, we proposed the AOMEB algorithm to maintain a $(\sqrt{2}+\varepsilon)$-coreset for
MEB in an append-only stream.
Then, based on AOMEB, we proposed two sliding-window algorithms, namely SWMEB and SWMEB+,
for coreset maintenance with constant approximation ratios.
We further generalized our proposed algorithms for kernelized MEBs.
Empirically, SWMEB and SWMEB+ improved the efficiency of the state-of-the-art
batch algorithm by up to four orders of magnitude while providing coresets with
rather small errors compared to the optimal ones.
For future work, we plan to explore the applications of coresets to various
data mining and machine learning problems in the streaming or sliding window model. 

\bibliographystyle{plainnat}
\bibliography{references}

\appendix
\section{Details on Experimental Setup}
In this section, we present the details of our experimental setup for reproducibility.

\subsection{Hardware Configuration}\label{appendix:hardware}
All the experiments are conducted on a server with the following specifications:
\begin{itemize}
    \item \textbf{CPU:} Intel(R) Xeon(R) E7-4820 v3 @ 1.90GHz
    \item \textbf{Memory:} 128GB (8$\times$16GB) RAM 2133MHz DDR4 memory
    \item \textbf{Hard Disk:} a 480GB SATA-III solid-state drive
\end{itemize}

\subsection{Software Environment}\label{appendix:software}
The server runs Ubuntu GNU/Linux 16.04.3 LTS 64-bit with kernel v4.11.0-rc2.
All the code is written in Java 8 only using the standard libraries.
No third-party software/library is required.
The version of JVM for compilation is Java HotSpot(TM) 64-Bit Server VM 18.3 (build 10.0.2+13).
In the experiments, each instance is limited to use a single thread for computation
to ensure the fairness of comparison.
In addition, we use a JVM option ``\texttt{-Xmx80000m}'' to restrict the maximum heap size
used by each instance. Note that the memory usage is much less than 80GB and
our purpose is to guarantee that the bottleneck is not in memory and I/O.

\subsection{Datasets}\label{appendix:datasets}
The statistics of the datasets used in our experiments are listed in Table~\ref{tbl:dataset}.
Here we briefly describe the real-world datasets we use and the preprocessing procedures.
\begin{itemize}
    \item \textsf{Census} contains a one percent sample of the Public Use Microdata Samples person records
    drawn from the full 1990 census sample.
    It is downloaded from UCI Machine Learning Repository.
    \item \textsf{CovType} is a dataset for predicting forest cover type from cartographic variables.
    It is available on UCI Machine Learning Repository and LIBSVM.
    We download the dataset from LIBSVM.
    In the preprocessing, we only retain the data with class label ``0''.
    \item \textsf{GIST} is an image collection retrieved from
    TEXMAX.
    \item \textsf{Gowalla} is a collection of user check-ins over the period of February 2009 -- October 2010
    on gowalla.com. It is downloaded from SNAP.
    In the preprocessing, we extract the latitude and longitude of each check-in
    as a two-dimensional point and dispose other attributes.
    \item \textsf{HIGGS} is a dataset for distinguishing between a signal process which produces Higgs bosons
    and a background process which does not. It is downloaded from UCI Machine Learning
    Repository.
    In the preprocessing, we only retain the data with class label ``1'' (i.e., signal process).
    \item \textsf{SIFT} is an image collection retrieved from
    TEXMAX.
\end{itemize}

The procedure of generating \textsf{Synthetic} has been introduced
in the main paper. We transform each dataset into a single file that stores the points in dense format:
each point is represented by one line in the file and different dimensions of the point are split by a single space.
After dataset preprocessing and format transformation, we shuffle all points of each dataset randomly.
It is guaranteed that a dataset must be processed by all algorithms in the same order
for the fairness of comparison.

\subsection{Implementation Issues}\label{appendix:impl}
The algorithms we compare in the experiments are as listed in Section~\ref{subsec:exp:setup}.
The implementations of our algorithms are available
at \url{https://github.com/yhwang1990/SW-MEB}.
Here we discuss the implementations of these algorithms.

We use the Java code published by the authors,
which is available at \url{https://github.com/hbf/miniball},
for the implementation of COMEB~\cite{DBLP:conf/esa/FischerGK03}.
All the other algorithms are implemented by ourselves.
First of all, the basic scheme of CoreMEB~\cite{DBLP:journals/comgeo/BadoiuC08} is
presented in Algorithm~\ref{alg:coreset:offline}. Our practical implementations
are based on two improved versions of CoreMEB,
i.e., Figure 2 in~\cite{DBLP:journals/cgf/LarssonK13} for Euclidean MEB
and Algorithm 4.1 in~\cite{DBLP:journals/siamjo/Yildirim08} for kernelized MEB.
They use the same scheme as shown in Algorithm~\ref{alg:coreset:offline}
but have lower computational costs and quicker convergence rates.
The implementation of SSMEB is based on Section 2 of~\cite{DBLP:conf/cccg/Zarrabi-ZadehC06},
which is extended to kernelized MEB according to Section 4.2 of~\cite{DBLP:conf/ijcai/RaiDV09}.
We implement the Blurred Ball Cover (BBC) algorithm according to
Section 2 of~\cite{DBLP:conf/soda/AgarwalS10}.
In addition, DyMEB is implemented based on Algorithms 1--3 in~\cite{DBLP:journals/comgeo/ChanP14}.
Finally, our proposed methods, i.e., AOMEB, SWMEB and SWMEB+, are implemented
according to Algorithms~\ref{alg:coreset:append:only}--\ref{alg:swmeb:plus} in this paper.

\subsection{Parameter Tuning}\label{appendix:parameter}
The procedure of parameter tuning is as follows.
Generally, there are three parameters used in our experiments:
$\varepsilon_1$ (or $\varepsilon$) in all algorithms except COMEB and SSMEB,
$\varepsilon_2$ in SWMEB and SWMEB+,
the partition size $L$ in SWMEB.

First of all, we choose appropriate $\varepsilon_1$
for Euclidean and kernelized MEBs respectively.
The parameter $\varepsilon_1$ determines the trade-off between coreset quality and efficiency.
We test the effect of $\varepsilon_1$ by validating on $[10^{-1}, 10^{-2},\ldots,10^{-6}]$.
For Euclidean MEB, the coreset quality hardly improves but the running time increases rapidly
when $\varepsilon_1 < 10^{-3}$.
Therefore, we set $\varepsilon_1 = 10^{-3}$ for Euclidean MEB.
Using the same method, we set $\varepsilon_1 = 10^{-4}$ for kernelized MEB.

After choosing appropriate $\varepsilon_1$'s, we further test the remaining parameters in SWMEB and SWMEB+.
Specifically, $\varepsilon_2$ is also selected from $[10^{-1}, 10^{-2},\ldots,10^{-6}]$.
In SWMEB, $\varepsilon_2$ affects the index construction on each partition.
When $\varepsilon_2 \leq 0.01$, SWMEB suffers from high overhead because of containing too many indices.
Thus, we use $\varepsilon_2=0.1$ for SWMEB. Additionally, to avoid the indices
being too sparse, we restrict the maximal distance between any neighboring indices
in the same partition to $\frac{L}{10}$.
In SWMEB+, $\varepsilon_2$ adjusts the number of indices in $X_t$.
Firstly, we observe that the coreset quality cannot be improved any more when $\varepsilon_2 < \frac{\varepsilon_1}{10}$.
Secondly, we scale $\varepsilon_2$ by a factor of $\lambda>1$ among indices,
i.e., $\varepsilon_2=\lambda^{i-1}\cdot\frac{\varepsilon_1}{10}$ for $x_i \in X_t$,
to reduce the index size without seriously affecting the quality.
We select $\lambda$ from $\{2,4,8,16\}$ and use $\lambda=4$ for SWMEB+ since the quality seriously degrades when $\lambda>4$.
Thirdly, we set $0.1$ as the upper bound of $\varepsilon_2$ to ensure the theoretical soundness.
To sum up, we use $\varepsilon_2=\min(4^{i-1} \cdot \frac{\varepsilon_1}{10}, 0.1)$ for each $x_i \in X_t$ in SWMEB+.
Finally, the partition size $L$ in SWMEB affects the balance between space and time complexity.
If $L$ is smaller, $W_t$ will be divided into more partitions, which leads to more indices in $X_t$,
but fewer points will be stored in the buffer $Q$, and vice versa.
We try to select $L$ from range $[\frac{N}{5},\frac{N}{10},\ldots]$.
The results show that SWMEB cannot scale to large datasets when $L < \frac{N}{10}$
for too many indices in $X_t$.
Therefore, the partition size $L$ in SWMEB is set to $\frac{N}{10}$.

\section{Additional Experimental Results}\label{appendix:results}
\begin{figure}
    \centering
    \begin{subfigure}{0.96\textwidth}
        \includegraphics[width=\textwidth]{legend-eps-converted-to.pdf}
    \end{subfigure}
    \vskip\baselineskip
    \vspace{-1em}
    \begin{subfigure}{0.3\textwidth}
        \includegraphics[width=\textwidth]{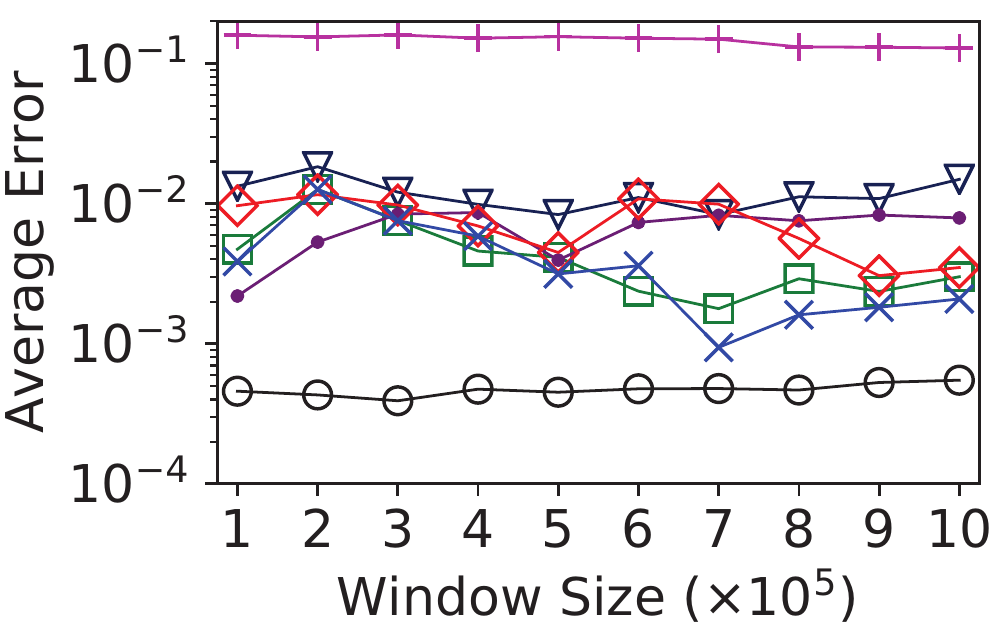}
    \end{subfigure}
    \begin{subfigure}{0.3\textwidth}
        \includegraphics[width=\textwidth]{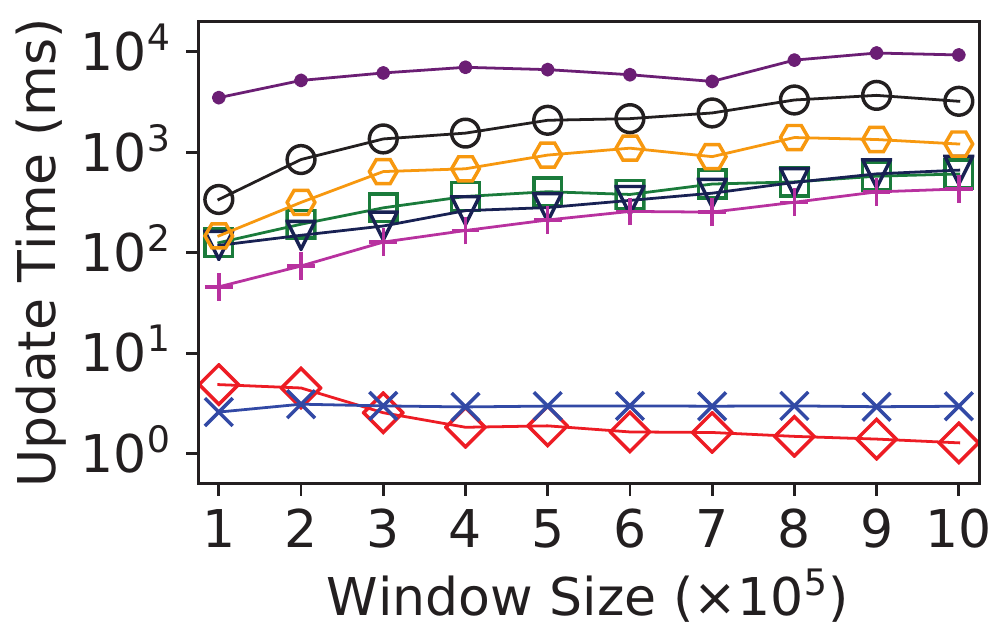}
    \end{subfigure}
    \begin{subfigure}{0.3\textwidth}
        \includegraphics[width=\textwidth]{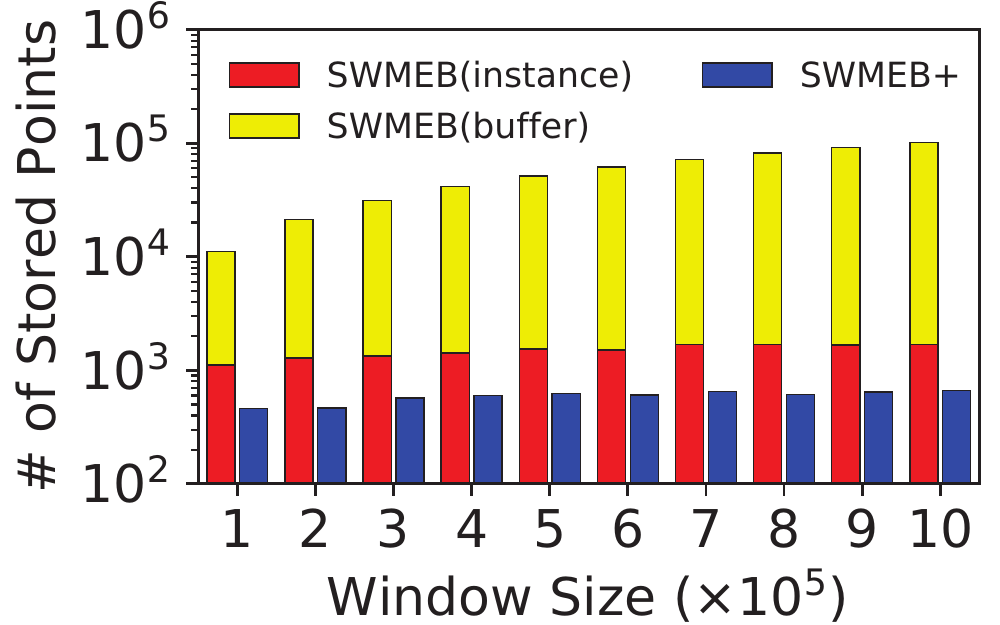}
    \end{subfigure}
    \caption{The performance for Euclidean MEB with varying $N$ on the HIGGS dataset.}
    \label{fig:Euclidean:N:Higgs}
\end{figure}

\begin{figure}
    \centering
    \begin{subfigure}{0.3\textwidth}
        \includegraphics[width=\textwidth]{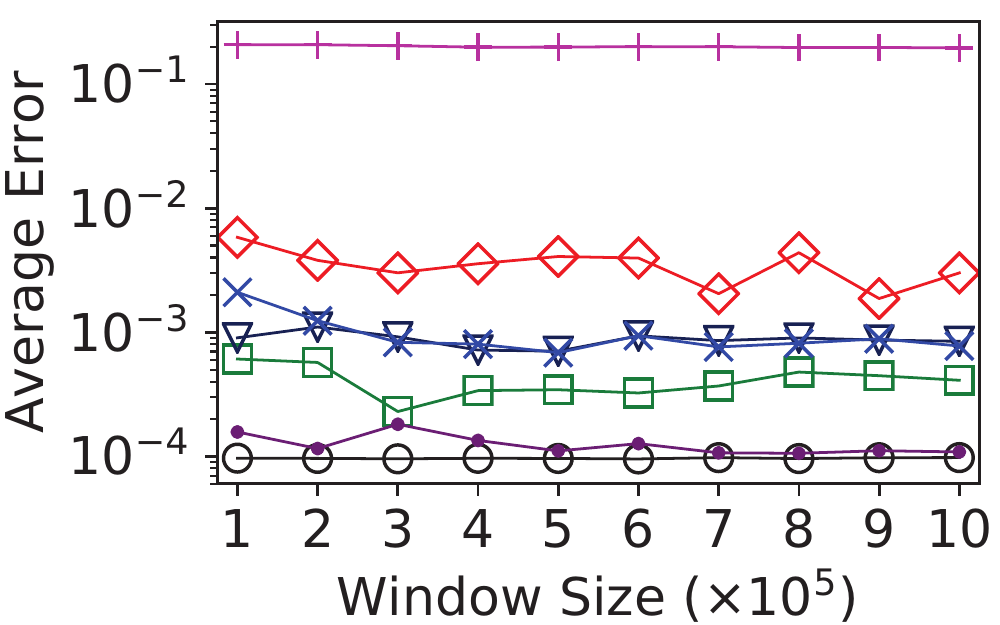}
    \end{subfigure}
    \begin{subfigure}{0.3\textwidth}
        \includegraphics[width=\textwidth]{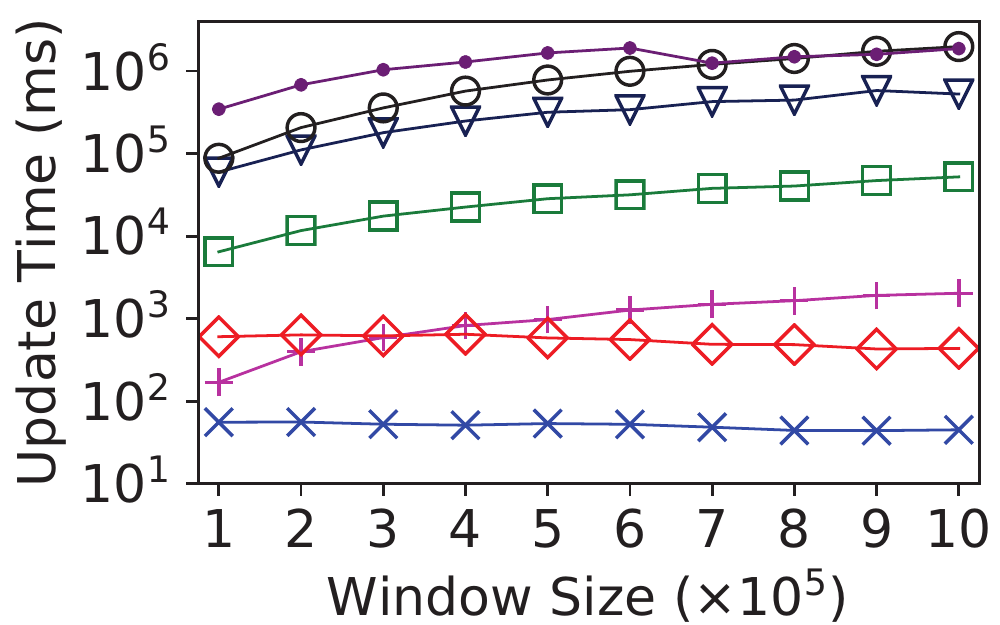}
    \end{subfigure}
    \begin{subfigure}{0.3\textwidth}
        \includegraphics[width=\textwidth]{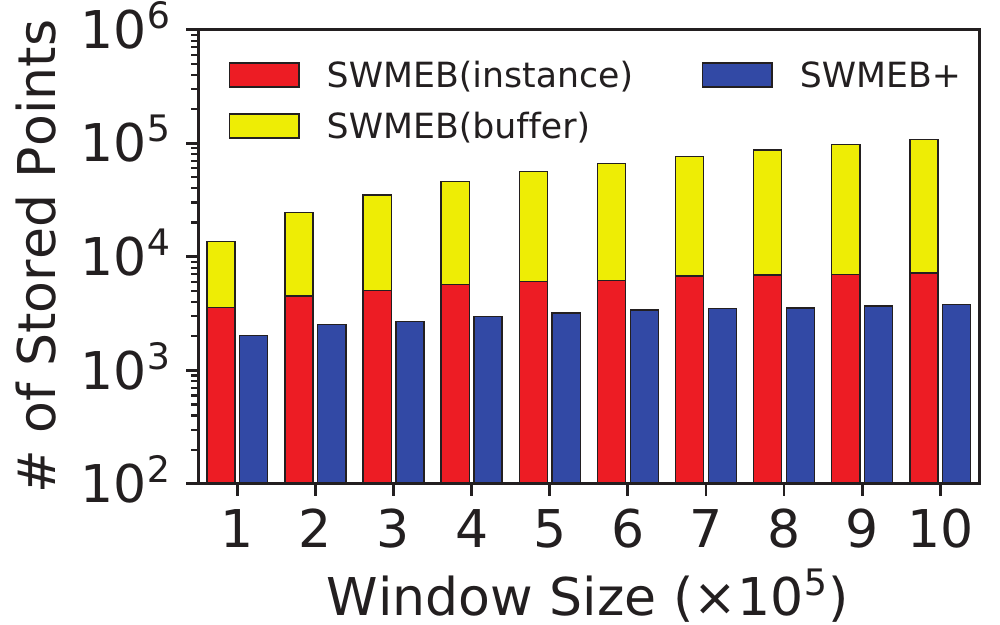}
    \end{subfigure}
    \caption{The performance for kernelized MEB with varying $N$ on the HIGGS dataset.}
    \label{fig:kernel:N:Higgs}
\end{figure}

\begin{figure}
    \centering
    \begin{subfigure}{0.3\textwidth}
        \includegraphics[width=\textwidth]{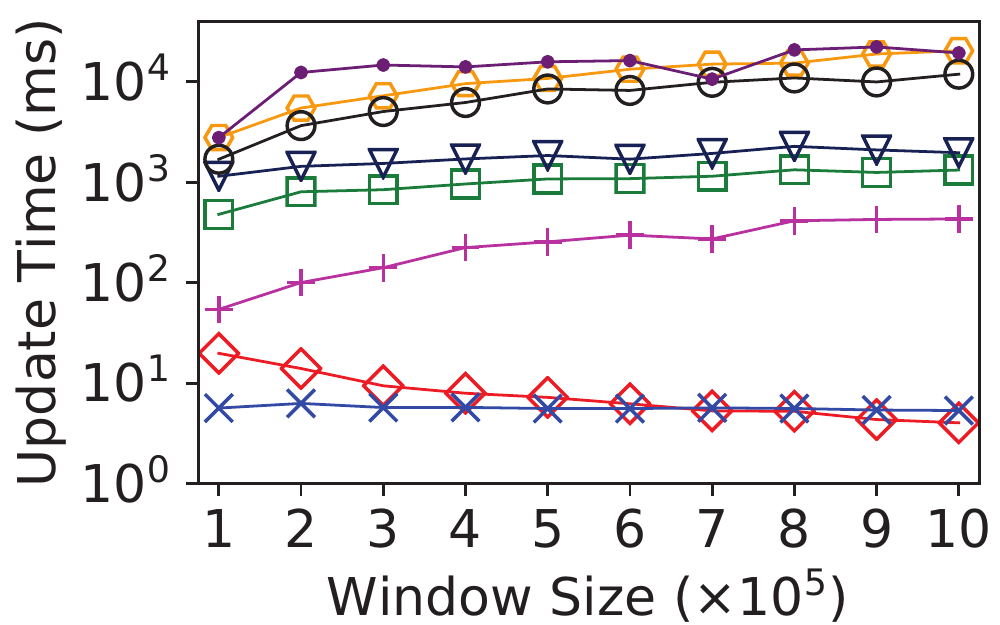}
    \end{subfigure}
    \begin{subfigure}{0.3\textwidth}
        \includegraphics[width=\textwidth]{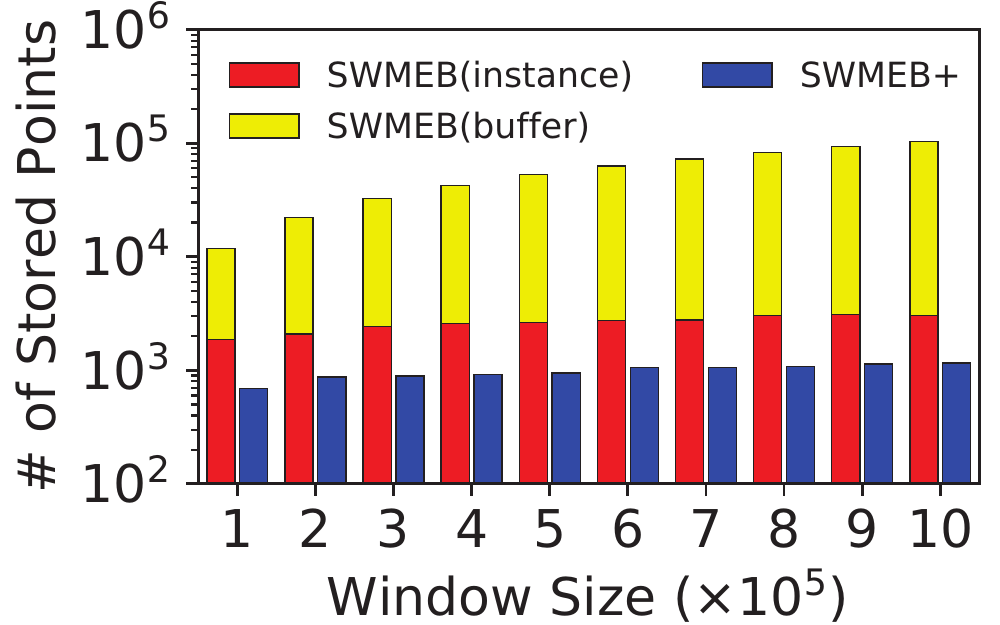}
    \end{subfigure}
    \caption{The performance for Euclidean MEB with varying $N$ on Synthetic dataset ($m=50$).}
    \label{fig:Euclidean:N:Synthetic}
\end{figure}

\begin{figure}
    \centering
    \begin{subfigure}{0.3\textwidth}
        \includegraphics[width=\textwidth]{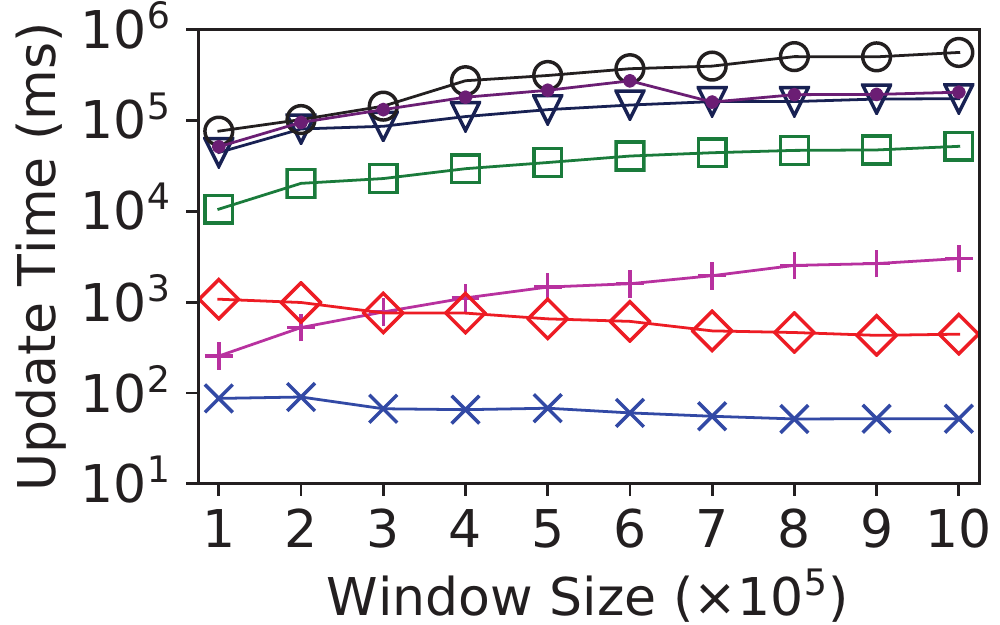}
    \end{subfigure}
    \begin{subfigure}{0.3\textwidth}
        \includegraphics[width=\textwidth]{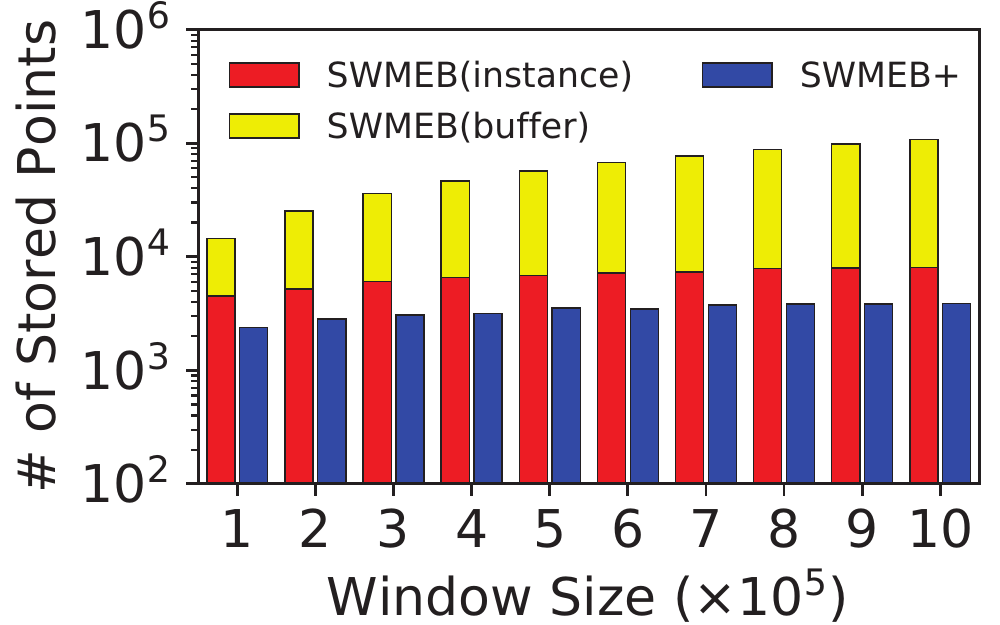}
    \end{subfigure}
    \caption{The performance for kernelized MEB with varying $N$ on Synthetic dataset ($m=50$).}
    \label{fig:N:Synthetic}
\end{figure}

\begin{figure}
    \centering
    \begin{subfigure}{0.3\textwidth}
        \includegraphics[width=\textwidth]{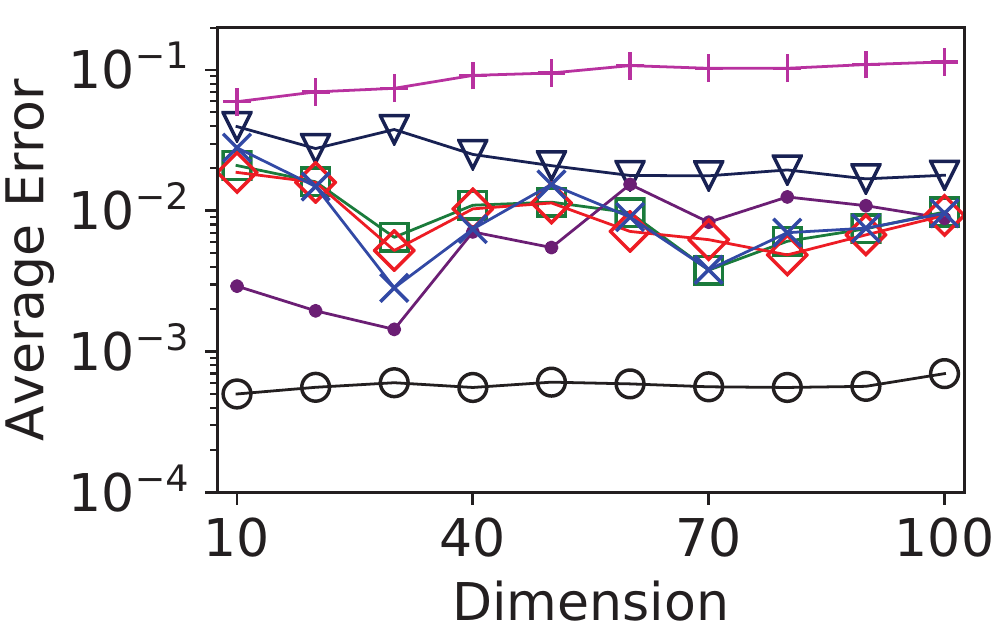}
    \end{subfigure}
    \begin{subfigure}{0.3\textwidth}
        \includegraphics[width=\textwidth]{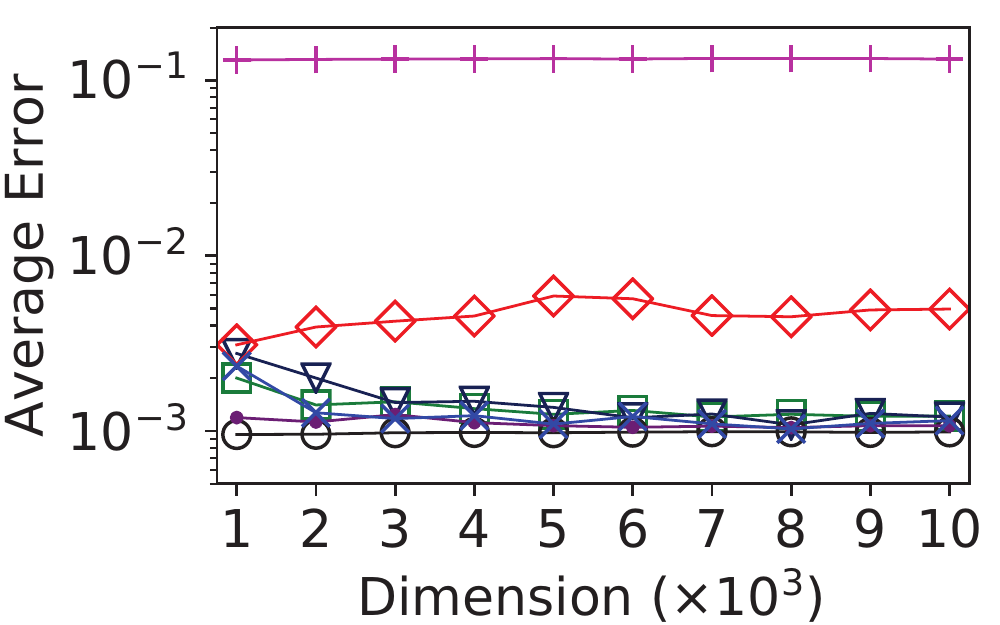}
    \end{subfigure}
    \caption{The average error for Euclidean MEB with varying $m$ on Synthetic dataset.}
    \label{fig:m:Synthetic:error}
\end{figure}

We present additional experimental results for scalability on different datasets
in Figures~\ref{fig:Euclidean:N:Higgs}--\ref{fig:m:Synthetic:error}.

\end{document}